\documentclass[journal]{IEEEtran}
\usepackage{graphicx, pstool}
\usepackage{amsmath}
\usepackage{amsfonts}
\usepackage{amssymb,bbm}
\usepackage{tabularx}
\usepackage{url}

\begin{document}

\title{  Achievable Rates  for the Fading Half-Duplex Single Relay Selection Network Using Buffer-Aided Relaying }
\author{Nikola~Zlatanov,~\IEEEmembership{Student Member,~IEEE,}   Vahid Jamali,~\IEEEmembership{Student Member,~IEEE,}~and~Robert~Schober,~\IEEEmembership{Fellow,~IEEE}
\thanks{Manuscript received August 26, 2014; revised January 18, 2015; accepted April 1, 2015. This paper has been presented in part at IEEE Globecom 2014, Austin, TX, December 2014.}
\thanks{N. Zlatanov is with the Department of Electrical and Computer Engineering, University of British Columbia, Vancouver, BC, V6T 1Z4,
Canada, E-mail: zlatanov@ece.ubc.ca}
\thanks{V. Jamali and R. Schober are with the Institute for Digital Communication,
Friedrich-Alexander University, Erlangen 91054, Germany (e-mail:
jamali@lnt.de; schober@lnt.de).}
}

\maketitle

\begin{abstract}
In the half-duplex  single relay selection network, comprised of a source, $M$ half-duplex relays, and a destination, only one relay is active  at any given time, i.e.,  only one relay receives or transmits, and the other relays are inactive, i.e., they do not  receive nor transmit. The capacity of this network, when all links are affected by independent  slow  time-continuous  fading and additive white Gaussian noise (AWGN), is still unknown, and  only achievable average rates have been reported in the literature so far. In this paper, we present new achievable average rates for this network which are larger than the best known average rates. These new average rates are achieved with a buffer-aided relaying protocol. Since the developed buffer-aided protocol    introduces unbounded delay,  we also devise a buffer-aided protocol which limits the delay at the expense of a decrease in rate. Moreover, we discuss the practical implementation of the proposed  buffer-aided relaying protocols and show that they do not require more resources for channel state information acquisition than  the existing relay selection protocols.
\end{abstract}


\IEEEpeerreviewmaketitle
 
\newtheorem{theorem}{Theorem}
\newtheorem{lemma}{Lemma}
\newtheorem{corollary}{Corollary}
\newtheorem{remark}{Remark}
\newtheorem{definition}{Definition}
\newtheorem{proposition}{Proposition}
 

\section{Introduction}

\IEEEPARstart{C}{ooperative} communication  has recently gained
 much attention due to its ability to increase the throughput and/or reliability of wireless networks. The basic idea behind cooperative communication is that each node can act as a relay and help the other nodes of the network to
forward their  information to their respective destination nodes. Because of the high complexity inherent to the investigation of general cooperative networks, and to get insight into the basic challenges and benefits of cooperative communication, researchers have mainly considered relatively simple cooperative networks. Although simple, these basic cooperative networks reveal the gains that can be accomplished by cooperation among  network nodes. Moreover, because of their simplicity, these basic cooperative networks  can be easily integrated into the current communication infrastructure.  One basic network which has shown great potential in terms of utility and performance is the
 half-duplex (HD) single relay selection network  proposed in \cite{Bletsas06}. In this network,  only one relay is active at any given time, i.e., one relay receives or transmits, and the other relays are inactive, i.e., they do not  receive nor transmit.  Because of the large achievable performance gains, this network has recently attracted considerable interest, see \nocite{Bletsas06,  5290303, 5165288, 4801494, 4657317, 4641952, 4524271,  Krikidis, ikhlef2012max, 5351695, 5397898} \cite{Bletsas06}-\cite{5397898} and  references therein. Although well investigated, the capacity of this network is still unknown when all links are affected by independent  slow  time-continuous fading and additive white Gaussian noise (AWGN). So far, only achievable average rates\footnote{ The ``average rate''  is also referred to as ``expected rate''   in  the literature.} have been reported in the literature, see \cite{5351695, 5397898}. In fact, to the best of the authors' knowledge, the achievable average rates in \cite{5351695} and \cite{5397898} are the largest  average rates reported in the literature for this network. These rates are based on the relay selection protocol in \cite{Bletsas06}, where, in each time slot,  the  relay with the strongest minimum source-to-relay and relay-to-destination channel is selected to forward the information from the source to the destination. In this paper, we will show that these rates can be surpassed.  In particular,  we develop a buffer-aided relaying  protocol which achieves average rates  which are significantly larger than the rates reported in \cite{5351695} and \cite{5397898}.  Since the proposed buffer-aided protocol  introduces unbounded delay,  we also devise a second buffer-aided protocol which limits the average delay at the expense of a decrease in rate.  Moreover, we show that the proposed  buffer-aided relaying protocols do not require more resources for channel state information (CSI) acquisition than  the existing relay selection protocols.

Buffer-aided HD relaying with adaptive switching between reception and transmission was proposed in \cite{6133900} for a simple three-node relay network without source-destination link. Later, buffer-aided  relaying  was further analyzed in \cite{BA-relaying-adaptive-rate} and \cite{BA-relaying-fixed-mixed-rate} for adaptive and fixed rate transmission, respectively. Buffer-aided relaying protocols were also proposed for two-way relaying in \cite{petar_two_way}, \cite{jamali2013adaptive}, the multihop relay network in \cite{6189806}, two source   and two destination pairs  sharing a single relay in \cite{6783948}, secure communication  for two-hop relaying and relay selection  in \cite{6847181} and \cite{6746659}, respectively, and amplify-and-forward relaying in \cite{6817619}.
For the considered relay selection network, relaying with buffers   was investigated in \cite{Krikidis} and \cite{ikhlef2012max}. However, the protocols in \cite{Krikidis} and \cite{ikhlef2012max} are limited to the case when all nodes transmit with fixed rates and all source-to-relay and relay-to-destination links undergo independent and identically distributed (i.i.d.) fading. These  protocols were developed for improving the outage probability performance of the network. In order to use the  protocols in \cite{Krikidis} and \cite{ikhlef2012max} as performance benchmarks, we modify them such that all nodes transmit with rates equal to their underlying channel capacities.  However, the modified protocols are still only applicable to the case when all links are affected by i.i.d.  fading and will cause data loss due to buffer overflow  for independent non-identically distributed (i.n.d.) fading. We note however  that this   drawback is not caused by our modifications since the phenomenon of buffer overflow  also occurs  for the original protocols in \cite{Krikidis} and \cite{ikhlef2012max} for  fixed rate transmission when the links of the network are  i.n.d.

This paper is organized as follows. In Section~\ref{sec_1}, we introduce the system model. In Section ~\ref{sec_2},  we present the proposed buffer-aided protocol  for transmission without delay constraints. In Section~\ref{sec_impl}, we discuss the implementation  of the proposed protocol. In Section~\ref{sec_delay-I}, we propose a protocol for delay-limited transmission. In  Section~\ref{sec_3},  we provide  numerical examples   comparing the achievable rates of the proposed protocols and the benchmark protocols. Finally, Section~\ref{sec_4}  concludes the paper.

\section{System Model} \label{sec_1}
In the following, we introduce the  system model of the considered relay network. Furthermore, as benchmark scheme, we briefly review the conventional  non-buffer-aided relay selection protocol in \cite{Bletsas06}.
\subsection{System Model}
\begin{figure}
\centering
\pstool[width=1\linewidth]{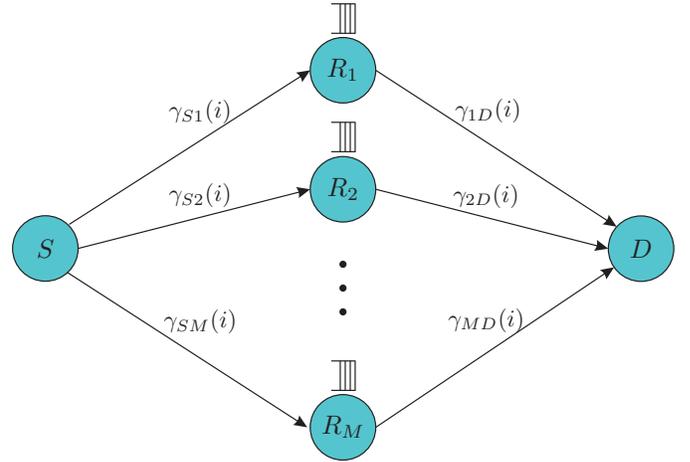}{
\psfrag{S}[c][c][1]{$S$}
\psfrag{D}[c][c][1]{$D$}
\psfrag{R1}[c][c][1]{$R_1$}
\psfrag{R2}[c][c][1]{$R_2$}
\psfrag{RM}[c][c][1]{$R_M$}
\psfrag{h0}[c][c][0.9]{$\gamma_{S1}(i)$}
\psfrag{h1}[c][c][0.9]{$\gamma_{S2}(i)$}
\psfrag{hM}[c][c][0.9]{$\gamma_{SM}(i)$}
\psfrag{g0}[c][c][0.9]{$\gamma_{1D}(i)$}
\psfrag{g1}[c][c][0.9]{$\gamma_{2D}(i)$}
\psfrag{gM}[c][c][0.9]{$\gamma_{MD}(i)$}
}
\caption{System model for buffer-aided relay selection.}
\label{sys_model}
\end{figure}

The HD relay selection network consists of a source $S$, $M$ HD decode-and-forward relays $R_k$, $k=1,...,M$, and a destination $D$, as shown in Fig.~\ref{sys_model}. The source transmits its information to the destination only through the relays, i.e., because of high attenuation there is no direct link between the source and the destination, and therefore, all the information that the destination receives is  first processed by   the relays. We assume that the transmission is performed in $N$ time slots, where $N\to\infty$.   The  relays in the network are HD nodes, i.e., they cannot transmit and receive at the same time. Furthermore, in each time slot, only one relay is active, i.e., it receives or transmits, and the other relays are inactive, i.e., they do not   receive nor transmit. 
Each relay is equipped with a buffer of unlimited size in which it stores the information that it receives from the source and from which it extracts the information that it transmits to the destination. We assume that all nodes transmit their codewords with constant power $P$ and that the noise at all receivers is independent AWGN with variance $\sigma_n^2$.  We assume transmission with capacity achieving codes. Hence, the transmitted codewords are Gaussian distributed,  comprised of  $n\to\infty$ symbols, and span one time slot.  Moreover,   we assume that each  source-to-relay   and relay-to-destination channel  is affected by independent  slow  time-continuous fading such that the fading remains constant during a single time slot and changes from one time slot to the next. We assume that the fading is an ergodic and stationary random process.   Let $|h_{Sk}(i)|^2$ and $|h_{kD}(i)|^2$ denote the squared amplitudes  of the complex channel gains of the  source-to-$k$-th-relay and $k$-th-relay-to-destination channels   in  the  $i$-th time slot, respectively, and let $\Omega_{Sk}=E\{|h_{Sk}(i)|^2\}$ and $\Omega_{kD}=E\{|h_{kD}(i)|^2\}$ denote their mean values, respectively, where $E\{\cdot\}$ denotes expectation. Then, the signal-to-noise ratios (SNRs) of the source-to-$k$-th-relay and $k$-th-relay-to-destination channels are given by
\begin{eqnarray}\label{err-1}
    \gamma_{Sk}(i)=\frac{P}{\sigma_n^2}|h_{Sk}(i)|^2\; \textrm{ and }\;  \gamma_{kD}(i)=\frac{P}{\sigma_n^2}|h_{kD}(i)|^2,
\end{eqnarray}
respectively. Furthermore, we denote the average SNRs of the  source-to-$k$-th-relay and $k$-th-relay-to-destination channels by $\bar \gamma_{Sk}=E\{\gamma_{Sk}(i)\}$ and $\bar \gamma_{kD}=E\{\gamma_{kD}(i)\}$, respectively. Using (\ref{err-1}),  the capacities of the source-to-$k$-th-relay and $k$-th-relay-to-destination channels in  the  $i$-th time slot, denoted by   $C_{S k}(i)$ and $C_{kD}(i)$, respectively, are given by
\begin{align}
  C_{S k}(i)&= \log_2\big(1+\gamma_{Sk}(i)\big)\label{eq_c_1} \\
 C_{k D}(i)&=\log_2\big(1+\gamma_{kD}(i)\big)\label{eq_c_2}.
\end{align}

\subsection{Conventional Relay Selection Protocol}\label{sec_II_conv}
 For comparison purpose, we briefly review the conventional  non-buffer-aided relay selection protocol  \cite{Bletsas06} and its corresponding achievable average rate \cite{5351695, 5397898}. 

 The conventional relay selection protocol  selects the relay $k$ with the maximum $\min\{C_{Sk}(i),C_{kD}(i)\}$ for forwarding the information from the source to the destination in the $i$-th time slot \cite{Bletsas06}.   The channel coding scheme adopted for conventional relaying is as follows. In the first half of time slot $i$, the source sends a codeword with rate $\min\{C_{Sk}(i), C_{kD}(i)\}$ to the $k$-th relay. The $k$-th relay can successfully decode the received codeword since the rate of the codeword is smaller than or equal to $C_{Sk}(i)$. Then, in the second half of time slot $i$, the relay re-encodes the decoded information and sends it to the destination with rate $\min\{C_{Sk}(i), C_{kD}(i)\}$. The destination can  successfully decode the received codeword since the rate of the codeword is smaller than or equal to $C_{kD}(i)$. Hence, the overall rate transmitted from source to destination  during  time slot $i$ is $\frac{1}{2}\min\{C_{Sk}(i), C_{kD}(i)\}$. Thereby, during $N\to\infty$ time slots, the average rate achieved with conventional relaying, denoted by $\bar R_{\rm conv}$, is obtained as \cite{5351695, 5397898}
\begin{eqnarray}\label{eq_r_conv}
  \bar R_{\rm conv}=\frac{1}{2} E\big\{\max_k\min\{C_{Sk}(i), C_{kD}(i)\}\big\} . 
\end{eqnarray}
In the following, we present the proposed buffer-aided protocols for the considered relay selection network and the corresponding achievable rates.

\section{Buffer-Aided Relaying Protocol without Delay Constraint}\label{sec_2}
 In this section, we develop a buffer-aided relaying protocol  without delay constraints which maximizes the  achievable  average rate for the considered network. To this end, we first introduce the instantaneous transmission rates at the nodes in each time slot, and then  derive the corresponding  achievable average rate. Next, we maximize the  achievable average rate  and derive  analytical expressions for the maximum average rate.

\subsection{Instantaneous Transmission Rates}
In the considered  HD single relay selection network, in a given time slot, only one relay is selected to receive or transmit, i.e., to be active. Without loss of generality, 
assume that the $k$-th relay  has been selected to be active in the $i$-th time slot\footnote{How exactly the active relay is selected is explained in Theorem 1.}. Then, if the active relay is selected to receive,  the source maps $n R_{Sk}(i)$ bits of information to a  Gaussian distributed codeword  comprised of $n\to\infty$ symbols, where each symbol  is generated independently according to a zero-mean complex circular-symmetric Gaussian distribution with variance $P$, and transmits this codeword to the selected relay. The rate  of this codeword   $R_{Sk}(i)$  is set as  
\begin{eqnarray}\label{eq_cc-1}
    R_{Sk}(i) =C_{S k}(i),
\end{eqnarray}
 where  $C_{Sk}(i)$ is the capacity of the source-to-$k$-th-relay  channel given in (\ref{eq_c_1}). As a result of (\ref{eq_cc-1}),   the active relay can successfully decode this codeword and stores the corresponding information in its buffer. Let $Q_k(i)$ denote the number of bits/symbol  in the buffer of the $k$-th relay at the end of  time slot $i$. Then, with this transmission,  $Q_k(i)$ increases as 
\begin{eqnarray}
     Q_k(i)=Q_k(i-1)+C_{S k}(i). 
\end{eqnarray}
 On the other hand, if the active relay  is selected to transmit, it extracts $n R_{kD}(i)$ bits of information from its buffer, maps it to a Gaussian distributed codeword    comprised of $n\to\infty$ symbols, where each symbol  is generated independently according to a zero-mean complex circular-symmetric Gaussian distribution with variance $P$, and transmits it to the destination.  The rate  of this codeword is $R_{kD}(i)$, which is set as  
\begin{eqnarray}\label{eq_cc-2}
    R_{kD}(i) =\min\{Q_k(i-1),C_{kD}(i)\},
\end{eqnarray}
where  $C_{kD}(i)$ is the capacity of the $k$-th-relay-to-destination channel given in (\ref{eq_c_1}).
 The minimum in the expression for rate $R_{kD}(i)$ is a consequence of the fact that the relay cannot transmit more information than what it has stored in its buffer, i.e., more than $Q_k(i-1)$.  The destination can successfully decode this codeword since $R_{kD}(i)\leq C_{kD}(i)$ holds, and stores the corresponding information.
When the active relay transmits,  $Q_k(i)$ decreases as  
\begin{eqnarray}
    Q_k(i)=Q_k(i-1)-R_{kD}(i).
\end{eqnarray}
In the following, we obtain the average rates of buffer-aided single-relay selection.

\subsection{ Average Transmission and Reception Rates}

In order to derive the average rates of  buffer-aided single-relay selection, we first have to model the reception and transmission of the $k$-th relay. To this end, we introduce 
  two  binary indicator variables   $r_k^{\rm R}(i)\in\{0,1\}$ and $r_k^{\rm T}(i)\in\{0,1\}$, which indicate whether, in the $i$-th time slot, the $k$-th relay     receives or transmits, respectively. More precisely, $r_k^{\rm R}(i)$ and $r_k^{\rm T}(i)$ are defined as
\begin{eqnarray}\label{eq_relay_R}
r_k^{\rm R}(i)\triangleq\left\{
\begin{array}{cl}
1 & \textrm{if the }  k\textrm{-th relay receives} \\
0 & \textrm{if the }  k \textrm{-th relay does not receive} ,
\end{array} 
\right.\;\\
r_k^{\rm T}(i)\triangleq\left\{
\begin{array}{cl}
1 & \textrm{if the }  k\textrm{-th relay transmits}  \\
0 & \textrm{if the }  k \textrm{-th relay does not transmit}  .
\end{array} 
\right.\label{eq_relay_T}
\end{eqnarray} 
Since exactly one relay is active in each time slot,  $r_k^{\rm R}(i)$ and $r_k^{\rm T}(i)$ must satisfy
\begin{eqnarray}\label{eq_c1}
    \sum_{k=1}^M [r_k^{\rm R}(i)+r_k^{\rm T}(i)] =1, \forall i.
\end{eqnarray}
Using $r_k^{\rm R}(i)$ and $r_k^{\rm T}(i)$, the average rates received at and transmitted by   the $k$-th relay, denoted by $\bar R_{Sk}$ and $\bar R_{kD}$, respectively, can be expressed as
\begin{align}
    \bar R_{Sk}&=\lim_{N\to\infty}\frac{1}{N}\sum_{i=1}^N r_k^{\rm R}(i)  R_{Sk}(i)\nonumber\\
&=\lim_{N\to\infty}\frac{1}{N}\sum_{i=1}^N r_k^{\rm R}(i) C_{Sk}(i),\label{eq_3s}\\
    \bar R_{kD}&=\lim_{N\to\infty}\frac{1}{N}\sum_{i=1}^Nr_k^{\rm T}(i) R_{kD}(i)\nonumber\\
&
=\lim_{N\to\infty}\frac{1}{N}\sum_{i=1}^N r_k^{\rm T}(i) \min\{Q_k(i-1),C_{kD}(i)\}.\label{eq_3} 
\end{align}
Using $\bar R_{kD}$, $\forall k$, the average rate received at the destination, denoted by $\bar R_{SD}$, can be expressed as
\begin{align}\label{eq_2}
    \bar R_{SD} &=\sum_{k=1}^M \bar R_{kD}\nonumber\\
&=\lim_{N\to\infty}\frac{1}{N}\sum_{i=1}^N \sum_{k=1}^M r_k^{\rm T}(i) \min\{Q_k(i-1),C_{kD}(i)\}. 
\end{align}

In the following, our goal is to maximize  $\bar R_{SD}$.

\subsection{ Maximization of the Average Rate}

In (\ref{eq_3s}) and (\ref{eq_3}), the only variables with a degree of freedom   are  $r_k^{\rm R}(i)$ and $r_k^{\rm T}(i)$, $\forall i,k$. Any choice of these variables  will provide an average  rate. However, in order for an average rate to be achievable, i.e., for data loss  not to occur, the buffers  at all relays must remain stable\footnote{By a stable buffer we mean that there is no information loss in the buffer and the information that enters the buffer eventually leaves the buffer, i.e., no information is trapped inside the buffer.}. Moreover, among all the achievable average  rates, there exists one rate which is the  largest.  In order to obtain the largest achievable average rate, we have to find the optimal values of  $r_k^{\rm R}(i)$ and $r_k^{\rm T}(i)$, $\forall i,k$, which maximize the  average rate in (\ref{eq_2}) when constraint (\ref{eq_c1}) holds and when the buffers at all relays are stable. To this end, we introduce the following useful lemma.
\begin{lemma}\label{lema_1}
The achievable average  rate is maximized when $r_k^{\rm R}(i)$ and $r_k^{\rm T}(i)$, $\forall i$, are chosen such that  the following condition is satisfied for all $k=1,...,M$  
\begin{eqnarray}\label{eq_k}
 \lim_{N\to\infty}\frac{1}{N}\sum_{i=1}^N r_k^{\rm R}(i) C_{Sk}(i)   = \lim_{N\to\infty}\frac{1}{N}\sum_{i=1}^N r_k^{\rm T}(i) C_{kD}(i).
\end{eqnarray}
Moreover, when (\ref{eq_k}) holds for the $k$-th relay, (\ref{eq_3}) simplifies to
\begin{eqnarray}\label{eq_ka}
     \bar R_{kD} 
=\lim_{N\to\infty}\frac{1}{N}\sum_{i=1}^N r_k^{\rm T}(i)  C_{kD}(i) ,
\end{eqnarray}
\end{lemma}
and   when (\ref{eq_k}) holds  $\forall k$ relays, (\ref{eq_2}) simplifies to
\begin{eqnarray}\label{eq_rate}
    \bar R_{SD} 
=\lim_{N\to\infty}\frac{1}{N}\sum_{i=1}^N \sum_{k=1}^M r_k^{\rm T}(i)  C_{kD}(i) .
\end{eqnarray}
\begin{IEEEproof}
Please refer to Appendix~\ref{app_lema_1}.
\end{IEEEproof}
With Lemma~\ref{lema_1}, we have reduced the search space for the maximum  achievable average rate to only those rates for which  (\ref{eq_k}) holds $\forall k$. Moreover, we have obtained an   expression for  $\bar R_{SD}$  which is independent of $Q_k(i)$, $\forall i,k$. Now, in order  to find the maximum  achievable average rate, we devise a maximization problem   for the  average rate, $\bar R_{SD}$, under the constraints given in (\ref{eq_k}) and   (\ref{eq_c1}). This maximization problem, for $N\to\infty$, is given by
\begin{eqnarray}\label{MPR1}
\begin{array}{ll}
 {\underset{r_k^{\rm R}(i),r_k^{\rm T}(i),\forall i,k}{\rm{Maximize: }}}&\frac{1}{N}\sum_{i=1}^N \sum_{k=1}^M r_k^{\rm T}(i) C_{k D}(i)\\
\vspace{1mm}
{\rm{Subject\;\; to: }} &{\rm C1:}\, \frac{1}{N}\sum_{i=1}^N  r_k^{\rm R}(i) C_{S k}(i) \\
&\qquad =\frac{1}{N}\sum_{i=1}^N  r_k^{\rm T}(i) C_{kD}(i),\; \forall k \\
\vspace{1mm}
  &{\rm C2:} \, r_k^{\rm R}(i)\in\{0,1\},\; \forall k,i \\
\vspace{1mm}
 &{\rm C3:} \, r_k^{\rm T}(i)\in\{0,1\},\; \forall k,i \\
\vspace{1mm}
&{\rm C4:} \, \sum_{k=1}^M [r_k^{\rm R}(i)+r_k^{\rm T}(i)] =1,\; \forall i. \\
\end{array}
\end{eqnarray}
In (\ref{MPR1}), the restrictions in  (\ref{eq_k}) and   (\ref{eq_c1}) are reflected in   constraints C1 and C4, respectively.  Fortunately, (\ref{MPR1}) can be solved analytically. The solution reveals how the  values of $r_k^{\rm R}(i)$ and $r_k^{\rm T}(i)$ are to be chosen optimally in each time slot $i$ such that the maximum  average rate of the buffer-aided protocol is achieved.  Before providing the solution to (\ref{MPR1}),  we first introduce some notations. 
Let  $\mu_k$,   $k=1,...,M$,  denote   constants which are independent of the time slot $i$ and the instantaneous CSI. The values of these constants depend on the   fading statistics and will be determined later, cf. Lemma~\ref{lema_2}.  Then, for a given time slot $i$, we multiply each $C_{Sk}(i)$ with $\mu_k$ and each  $C_{kD}(i)$ with $(1-\mu_k)$,  and collect these products  in  set $\mathcal{A}(i)$. Hence, $\mathcal{A}(i) $ is given by
\begin{align}\label{eq_A}
  & \mathcal{A}(i)=  \big\{ \mu_1 C_{S1}(i), \mu_2 C_{S2}(i),...,\mu_M C_{SM}(i), \nonumber\\
&  (1-\mu_1) C_{1D}(i),(1-\mu_2)  C_{2D}(i),..., (1-\mu_M)  C_{MD}(i) \big\} .  
\end{align}
 We are now ready to present the solution to (\ref{MPR1}) in the following theorem, which represents the proposed protocol for transmission without delay constraints.
\begin{theorem}\label{theo_2}
The optimal  values of $r_k^{\rm T}(i)$ and $r_k^{\rm R}(i)$, $\forall k,i$ which maximize the  achievable average  rate of the proposed protocol are given by
\begin{eqnarray}\label{eq_1_n}
\left\{\hspace{-2mm}
\begin{array}{ll}
\vspace{1mm}
 r_k^{\rm T}(i)=1  & \textrm{if }\;  (1-\mu_k)  C_{kD}(i) =\max \mathcal{A}(i)  \\
\vspace{1mm} 
  r_k^{\rm R}(i)=1 & \textrm{if }\;  \mu_k  C_{Sk }(i) =\max \mathcal{A}(i)\\
r_k^{\rm T}(i)=r_k^{\rm R}(i)=0  &  \textrm{otherwise}  ,  
\end{array}
\right.
\end{eqnarray}
 where the $\mu_k$,   $\forall k$, are chosen such that constraint C1 in (\ref{MPR1}) is satisfied $\forall k$. The maximum  achievable  average rate of the proposed protocol  is given by  (\ref{eq_rate}) when $r_k^{\rm R}(i)$ and $r_k^{\rm T}(i)$ are set as in  (\ref{eq_1_n}), $\forall i,k$.
\end{theorem}
\begin{IEEEproof} 
Please see Appendix~\ref{app_a}.
\end{IEEEproof}
\begin{remark}\label{remark_1}
Theorem~\ref{theo_2} reveals that the optimal values of $r_k^{\rm R}(i)$ and $r_k^{\rm T}(i)$ depend only on the instantaneous CSI of the $i$-th time slot, and are independent of the instantaneous CSIs of past and future time slots.
\end{remark}

\subsection{Analytical Characterization  of the Maximum Achievable Rate}

By inserting (\ref{eq_1_n}) into (\ref{eq_2}), we obtain the maximum  achievable rate of the proposed protocol  as an average over  $N\to\infty$  time slots, which may not be  convenient from an  analytical point of view. Furthermore,   Theorem~\ref{theo_2} does not provide an expression  for  obtaining  constants $\mu_k$, $\forall k$. 
In order to obtain useful analytical expressions for  the maximum  achievable average rate and  constants $\mu_k$, $\forall k$, we exploit the assumed ergodicity and stationarity of the fading, and  write  (\ref{eq_k}) (i.e., constraint C1 in (\ref{MPR1}))  and  (\ref{eq_rate})   equivalently as 
\begin{align}
\left\{
\begin{array}{lll}
 E\{ \log_2(1+\Gamma_{S1}(i))\} &=& E\{\log_2(1+\Gamma_{1D}(i))\} \\
 E\{ \log_2(1+\Gamma_{S2}(i))\} &=& E\{\log_2(1+\Gamma_{2D}(i))\} \\
&\vdots&\\
E\{ \log_2(1+\Gamma_{SM}(i))\} &=& E\{\log_2(1+\Gamma_{MD}(i))\}
\end{array}
\right.  \label{a_11c}  
\end{align}
 and  
\begin{eqnarray}
\bar R_{SD}&=&\sum_{k=1}^M  E\{\log_2(1+\Gamma_{kD}(i))\}\label{a_11b} ,
\end{eqnarray}
respectively, where $\Gamma_{Sk}(i) = r^{
\rm R}_k(i)\gamma_{Sk}(i)$ and $\Gamma_{kD}(i) = r^{\rm T}_k(i)\gamma_{kD}(i)$,  with $r_k^{\rm R}(i)$ and $r_k^{\rm T}(i)$  as in  (\ref{eq_1_n}).
In the following two lemmas, we provide simplified expressions for the maximum average rate $\bar R_{SD}$ and constants $\mu_k$, $\forall k$. Thereby, we drop   index $i$  since,  due to the stationarity  and ergodicity  of the fading, the statistics of $\Gamma_{Sk}(i)$ and $\Gamma_{kD}(i)$ are independent of $i$. 
\begin{lemma}\label{lema_2} 
 The optimal values of $\mu_k$, $k=1,...,M$, denoted by   $\mu_k^*$,  which maximize  $\bar R_{SD}$, can be obtained by solving\footnote{A system of nonlinear equations can be solved e.g. by algorithms based on Newton's method \cite{deuflhard2011newton}.} the following system  of $M$ equations
\begin{align}
\left\{\hspace{-2mm}
\begin{array}{lll}
 \int_{0}^\infty  \log_2(1+x) f_{\Gamma_{S1}}(x) dx  = \int_{0}^\infty  \log_2(1+x) f_{\Gamma_{1D}}(x) dx \\
\int_{0}^\infty  \log_2(1+x) f_{\Gamma_{S2}}(x) dx   = \int_{0}^\infty   \log_2(1+x) f_{\Gamma_{2D}}(x) dx \\
\hspace{40mm} \vdots \\
\int_{0}^\infty \log_2(1+x) f_{\Gamma_{SM}}(x) dx  =  \int_{0}^\infty  \log_2(1+x) f_{\Gamma_{MD}}(x) dx,
\end{array}
\right.  \label{a_11ca}
\end{align}
where, for $x>0$, 
\begin{align} 
& f_{\Gamma_{Sk}}(x) = f_{\gamma_{Sk}}(x)  F_{\gamma_{kD}}\left((1+x)^{\frac{\mu_k}{1-\mu_k}}-1\right)
\nonumber\\
& \qquad\times  \prod_{\underset{j\neq k}{j=1}}^M 
F_{\gamma_{Sj}}\left((1+x)^{\frac{\mu_k}{\mu_j}}-1\right)
F_{\gamma_{jD}}\left((1+x)^{\frac{\mu_k}{1-\mu_j}}-1\right),  \label{eq_d1a}
\\
&f_{\Gamma_{kD}}(x)=  f_{\gamma_{kD}}(x)  F_{\gamma_{Sk}}\left((1+x)^{\frac{1-\mu_k}{\mu_k}}-1\right)
\nonumber\\
& \times  \prod_{\underset{j\neq k}{j=1}}^M 
F_{\gamma_{Sj}}\left((1+x)^{\frac{1-\mu_k}{\mu_j}}-1\right)
F_{\gamma_{jD}}\left((1+x)^{\frac{1-\mu_k}{1-\mu_j}}-1\right). \label{eq_d1b}
\end{align}
Here,  $f_{\gamma_{\alpha}}(x)$ and $F_{\gamma_{\alpha}}(x)$ denote the probability density function (PDF) and cumulative distribution function (CDF) of $\gamma_{\alpha}$, $\alpha\in\{Sk,kD\}$, respectively.  Furthermore, if the fading on all  source-to-relay  and relay-to-destination  links is   i.i.d., the solution  to (\ref{a_11ca}) is $\mu_k^*=1/2$, $\forall k$.
\end{lemma}
\begin{IEEEproof}
Please refer to Appendix~\ref{app_pdf}.
\end{IEEEproof}
\begin{remark}\label{remark_2}
For i.i.d. links, since $\mu_k^*=1/2$, $\forall k$, the proposed protocol, given by  (\ref{eq_1_n}), always selects  the link with the largest instantaneous  channel gain among all $2M$ available links for transmission. Hence, for i.i.d. links this protocol becomes identical to the protocol   proposed in \cite{Krikidis}. However, for  i.n.d. links, the protocol in  \cite{Krikidis} will cause data loss due to buffer overflow. In particular, applying the protocols in \cite{Krikidis} and \cite{ikhlef2012max}, the buffers at relays with $\Omega_{Sk}>\Omega_{kD}$  suffer from overflow and receive more information than they can transmit. Hence, a fraction of the source's data is trapped inside the relay buffers and does not reach the destination, i.e., data loss occurs. On the other hand, our proposed protocol is applicable for all fading statistics.
\end{remark}
\begin{lemma}\label{lema_3}
The maximum  achievable average rate of the protocol in Theorem 1  is given by
    \begin{eqnarray}\label{eq_d2}
   \bar R_{SD}=  \sum_{k=1}^M \int_{0}^\infty \log_2(1+x) f_{\Gamma_{kD}}^*(x) dx,
\end{eqnarray}
where $f_{\Gamma_{kD}}^*(x)$ is obtained by inserting $\mu_k=\mu_k^*$  found using Lemma~\ref{lema_2} into $f_{\Gamma_{kD}}(x)$ given by (\ref{eq_d1b}).
For i.i.d. fading on all links, i.e., when $f_{\gamma_{Sk}}(x)=f_{\gamma_{kD}}(x)=f_{\gamma}(x)$, $\forall k$, and  $F_{\gamma_{Sk}}(x)=F_{\gamma_{kD}}(x)=F_{\gamma}(x)$, $\forall k$, (\ref{eq_d2}) simplifies to
\begin{eqnarray}\label{eq_d2a}
    \bar R_{SD}= M \int_{0}^\infty \log_2(1+x)  f_{\gamma}(x) \left(F_{\gamma}(x)\right)^{2M-1} dx.
\end{eqnarray}
\end{lemma}
\begin{IEEEproof}
  Let us  insert the optimal $\mu_k^*$, $\forall k$, found from Lemma~\ref{lema_2},  into $f_{\Gamma_{kD}}(x)$  given in (\ref{eq_d1b})  and denote  it  by $f_{\Gamma_{kD}}^*(x)$. 
Eq. (\ref{eq_d2}) is obtained by inserting $f_{\Gamma_{kD}}^*(x)$ into  (\ref{a_11b}), whereas  (\ref{eq_d2a}) is obtained by inserting $\mu^*=1/2$ into (\ref{eq_d2}) and simplifying the resulting expression.  
\end{IEEEproof}
To get more insight, in the following we investigate the case of i.i.d. Rayleigh fading.

\subsection{Special Case: I.i.d. Rayleigh Fading}
  In the following, we simplify the expression for the maximum average rate in (\ref{eq_d2a}) for i.i.d.  Rayleigh fading.
 
The expression $f_{\gamma_{\max}}(x)=2 M f_{\gamma}(x) \left(F_{\gamma}(x)\right)^{2M-1}$ in (\ref{eq_d2a}) can be interpreted as the distribution of the largest random variable (RV) among $2M$ i.i.d. RVs with distributions $f_{\gamma_{Sk}}(x)=f_{\gamma_{kD}}(x)=f_{\gamma}(x)$, $\forall k$,  see \cite{david1970order}.
 For i.i.d. Rayleigh fading, i.e., when  $f_{\gamma_{Sk}}(x)=f_{\gamma_{kD}}(x)=e^{-x/\bar \gamma}/\bar \gamma $, $\forall k$, where $\bar \gamma$ is the average SNR  of all source-to-relay and relay-to-destination links,  $f_{\gamma_{\max}}(x)$ is given as \cite{david1970order}
\begin{align}\label{eq_A3} 
    f_{\gamma_{\max}}(x)=2M \hspace{-1mm}  \sum_{k=0}^{2M-1} \hspace{-1mm} (-1)^k 
{2M-1\choose k} \frac{1}{\bar \gamma}\exp\left(\hspace{-1mm} -\frac{x}{\bar \gamma}(k+1) \right) .
\end{align}
Inserting (\ref{eq_A3}) into (\ref{eq_d2a}) and integrating, we obtain the  average rate as
\begin{align}
    \bar R_{SD}&= M \sum_{k=0}^{2M-1}  {2M-1\choose k} \frac{(-1)^k}{(1+k)\ln(2)} \exp\left(\frac{1+k}{\bar\gamma}\right)\nonumber\\
 &\times E_1\left(\frac{1+k}{\bar \gamma}\right), \label{eq_A4} 
\end{align}
where $E_1(\cdot)$ is the first order exponential integral function defined as $E_1(x)=\int_{1}^\infty e^{-xt}/(t) dt$. On  the other hand, for the same case, i.e., for i.i.d. Rayleigh fading on all links,  the achievable rate for conventional relay selection  given in (\ref{eq_r_conv})  can be written equivalently as  \cite[Eq. (26)]{dio_eh} 
\begin{align}\label{eq_A5} 
    \bar R_{\rm conv} &= \frac{M}{2} \sum_{k=0}^{M-1}  {M-1\choose k} \frac{(-1)^k}{(1+k)\ln(2)} \exp\left(\frac{2(1+k)}{\bar\gamma}\right) \nonumber\\
&\times E_1\left(\frac{2(1+k)}{\bar\gamma}\right).
\end{align}
In order to gain further insight, expressions  (\ref{eq_A4}) and  (\ref{eq_A5}) can be further simplified for low and high SNRs using the following first order Taylor approximations
\begin{align}
    & \exp(c/\bar \gamma)E_1\left(c/\bar \gamma \right)=\frac{c}{\bar\gamma},\textrm{ as }\bar\gamma\to 0,  \label{ap1}\\
&\exp(c/\bar \gamma)E_1\left(c  /\bar \gamma\right)=-K_{EM} - \ln(c)+ \ln(\bar\gamma) ,\textrm{ as }\bar\gamma\to \infty \label{ap2},
\end{align}
where $K_{EM}$ is the  Euler-Mascheroni constant and its value is $K_{EM}\approx 0.577$. 
\subsubsection{Low SNR}
Using (\ref{ap1}), the rates in (\ref{eq_A4}) and (\ref{eq_A5}) can be approximated as
\begin{align}
   & \bar R_{SD}\to \frac{\bar \gamma}{2 \ln(2) }\sum_{k=1}^{2M}\frac{1}{k}, \textrm{ as }\bar \gamma\to 0\label{eq_d2b} \\
&\bar R_{\rm conv}\to \frac{\bar \gamma}{4 \ln(2)   }\sum_{k=1}^{M}\frac{1}{k}, \textrm{ as }\bar \gamma\to 0.\label{eq_d2c} 
\end{align}
Dividing (\ref{eq_d2b}) by (\ref{eq_d2c}), we obtain the following ratio
\begin{equation}\label{eq_d2d} 
    \frac{\bar R_{SD}}{\bar R_{\rm conv}} =2\frac{\sum_{k=1}^{2M}\frac{1}{k}}{\sum_{k=1}^{M}\frac{1}{k}}.
\end{equation}
For $M=1$ and $M\to\infty$,  the ratio in (\ref{eq_d2d})  is equal to $3$ and $2$, respectively, which constitute  the upper and lower bounds of (\ref{eq_d2d}) for $1\leq M\leq \infty$. Hence, for low SNRs, the average rate of the proposed buffer-aided relay selection protocol is $2$ to $3$ times higher than the rate of conventional relay selection.
 \subsubsection{High SNR} On the other hand, using (\ref{ap2}), the rates in (\ref{eq_A4}) and (\ref{eq_A5}) can be approximated as
\begin{align}
   &  \bar R_{SD} \to \frac{\bar\gamma-K_{EM}}{2\ln(2)} \nonumber\\
&- M \sum_{k=0}^{2M-1}  {2M-1\choose k} \frac{(-1)^k\log_2(1+k)}{(1+k)} , \textrm{ as }\bar \gamma \to \infty, \label{eq_d2b1} 
\\
 &\bar R_{\rm conv} \to \frac{\bar\gamma-K_{EM}}{2\ln(2)} \nonumber\\
&\quad - \frac{M}{2} \sum_{k=0}^{M-1}  {M-1\choose k} \frac{(-1)^k\log_2(1+k)}{(1+k)} - \frac{1}{2}  , \textrm{ as }\bar \gamma \to \infty  .\label{eq_d2c1} 
\end{align}
Subtracting (\ref{eq_d2c1}) from (\ref{eq_d2b1}), we obtain 
\begin{align} 
   \bar R_{SD}-\bar R_{\rm conv}&=\frac{1}{2} + \frac{M}{2} \sum_{k=0}^{M-1}  {M-1\choose k} \frac{(-1)^k\log_2(1+k)}{(1+k)} \nonumber\\
&-M \sum_{k=0}^{2M-1}  {2M-1\choose k} \frac{(-1)^k\log_2(1+k)}{(1+k)}.\label{eq_d2da}
\end{align}
For $M=1$ and $M\to\infty$,  the expression in (\ref{eq_d2da})  evaluates to $1$ and $1/2$, respectively, which constitute the upper and lower bounds of (\ref{eq_d2da}) for $1\leq M\leq \infty$. Hence, for high SNRs, the average rate of the proposed buffer-aided relay selection protocol is between $1$ and $1/2$ bits/symb  larger than the rate of conventional relay selection.

 In the following, we discuss the implementation of the proposed buffer-aided HD relay selection protocol.

\section{Implementation of the Proposed Buffer-Aided Protocol}\label{sec_impl}
In this section, we discuss the implementation of the  protocol proposed in Theorem~\ref{theo_2}. The   proposed protocol can be implemented in a centralized or in a distributed manner. A centralized implementation  assumes  a central node which selects the active relay in each time slot and decides whether it should  receive or transmit. On the other hand, in the distributed implementation,   there is  no central node and the relays themselves negotiate which relay should  be active  in each time slot.   In the following, we discuss  both implementations.

\subsection{Centralized Implementation}\label{sec_c_i}
For the  centralized implementation, we assume that the destination is the central node. Hence,  in each time slot, the destination has to obtain  the CSI of  all links. To this end, at the beginning of each time slot, the source transmits pilot symbols from which all relays acquire their respective source-to-relay   CSIs. Then, each relay broadcasts orthogonal pilots, from which the source and destination learn all source-to-relay and relay-to-destination CSIs, respectively. Next, each relay feedsback\footnote{This feedback can also be done using pilots. In particular, since the destination already knows the channel between each relay and itself, each relay can broadcast pilots whose amplitude is equal to the channel gain  of the channel from the source to the selected relay.}   the CSI  of its respective source-to-relay channel to the destination. With the acquired CSI, the destination  computes $C_{Sk}(i)$ and $C_{kD}(i)$, $\forall k$. In order to select the active relay according to the protocol  in Theorem~\ref{theo_2}, the destination has to  construct  set $\mathcal{A}(i)$, given by (\ref{eq_A}). This requires the computation of the constants $\mu_k$, $\forall k$. These constants can be computed using Lemma~\ref{lema_2}, but this requires knowledge of the PDFs of the fading gains of all links before the start of transmission. Such a priori knowledge  may not be available  in practice.  In this case,  the destination has to  estimate $\mu_k$, $\forall k$, in real-time  using only the   CSI knowledge until time slot $i$. Since $\mu_k$, $\forall k$, are actually Lagrange multipliers obtained by solving the linear optimization problem in (\ref{eq:17a}), an accurate estimate of $\mu_k$, $\forall k$,  can be obtained using the gradient descent method  \cite{Boyd_CO}. In particular,
using $C_{Sk}(i)$ and $C_{kD}(i)$,  the destination recursively computes an estimate of $\mu_k$, denoted by $\mu_k^e(i)$,  as
\begin{eqnarray}\label{eq_f1}
     \mu_k^e(i)=\mu_k^e(i-1)+\delta_k(i) (\bar 
  R_{kD}^e(i-1)-  \bar R_{Sk}^e(i-1)),
\end{eqnarray}
where $\bar R_{Sk}^e(i-1)$  and $\bar R_{kD}^e(i-1)$ are real-time estimates of  $\bar R_{Sk}$  and $\bar R_{kD}$, respectively, computed for  $i\geq 2$ as
\begin{align}
    \bar R_{Sk}^e(i-1)&= \frac{i-2}{i-1} \bar R_{Sk}^e(i-2)+\frac{r_k^{\rm R}(i-1)}{i-1}C_{Sk}(i-1) ,\label{eq_f2a}\\
\bar R_{kD}^e(i-1)&= \frac{i-2}{i-1} \bar R_{kD}^e(i-2)+\frac{r_k^{\rm T}(i-1)}{i-1}C_{kD}(i-1) ,\label{eq_f2b}
\end{align}
where  $R_{Sk}^e(0)$ and $R_{kD}^e(0)$ are set to zero $\forall k$. 
In (\ref{eq_f1}),  $\delta_k(i)$ is an adaptive step size   which controls the speed of convergence of $\mu_k^e(i)$ to $\mu_k$. In particular, the step size $\delta_k(i)$  is  some properly chosen monotonically decaying function of $i$  with $\delta_k(1)<1$, see \cite{Boyd_CO} for more details.

Once the destination has $C_{Sk}(i)$, $C_{kD}(i)$, and $\mu_k^e(i)$, $\forall k$, it constructs the set $\mathcal{A}(i)$, and selects the active relay according to Theorem~\ref{theo_2}.  The destination also has to keep track of the queue length in the buffers at each relay in each time slot. To this end,  using  $C_{Sk}(i)$, $C_{kD}(i)$, $r_k^{\rm R}(i)$, and $r_k^{\rm T}(i)$, $\forall k$, the destination computes the queue length in the buffers at each relay using the following formula
\begin{align}\label{eq_qqq1}
    Q_k(i)&=Q_k(i-1)+ r_k^{\rm R}(i) C_{Sk}(i)\nonumber\\
& - r_k^{\rm T}(i) \min\{Q_k(i-1),C_{kD}(i)\}.
\end{align}
  Then, the destination broadcasts a control message to the relays which contains information regarding which relay is selected  and whether it will receive or transmit. If the selected relay is scheduled to transmit, it extracts information bits from its buffer, maps them to a codeword, and transmits the codeword to the destination    with rate $R_{kD}(i)=\min\{Q_k(i-1),C_{kD}(i)\}$. 
 Otherwise, if  the selected relay is scheduled to receive, it sends a control message to the source which  informs the source which relay is selected. Then, the source transmits  the information codeword intended for  the selected relay with rate $R_{Sk}(i)= C_{Sk}(i)$.

 The destination  may receive the information bits in an order which is different from that in which they were transmitted by the source. However, using the acquired CSI, the destination can keep track of the amount of information  received and transmitted by each relay in each time slot. This information is sufficient for the destination to perform successful reordering of the received information bits.

\subsection{Distributed Implementation}
 
We now outline the distributed implementation of the proposed protocol using timers, similar to the scheme  in \cite{Bletsas06}.

At the beginning of time slot $i$,  source and destination transmit pilots in successive pilot time slots. This enables the relays to acquire the CSI of their respective source-to-relay and relay-to-destination channels, respectively. Using the acquired  CSI, the $k$-th relay computes $C_{Sk}(i)$ and $C_{kD}(i)$.  Next,
using $C_{Sk}(i)$ and $C_{kD}(i)$,  the $k$-th relay computes the estimate of $\mu_k$,  $\mu_k^e(i)$, using (\ref{eq_f1}), (\ref{eq_f2a}), and (\ref{eq_f2b}).  Using $C_{Sk}(i)$, $C_{kD}(i)$,  and $\mu_k^e(i)$, the $k$-th relay turns on a timer proportional to $1/\max\{\mu_k^e(i) C_{Sk}(i), (1-\mu_k^e(i)) C_{kD}(i)\}$. This procedure is performed by all $M$ relays. If 
$$\max\{\mu_k^e(i) C_{Sk}(i), (1-\mu_k^e(i)) C_{kD}(i)\}=\mu_k^e(i) C_{Sk}(i)$$
 and 
$$\max\{\mu_k^e(i) C_{Sk}(i), (1-\mu_k^e(i)) C_{kD}(i)\}=(1-\mu_k^e(i)) C_{kD}(i),$$ the $k$-th relay knows that if it is selected, then it will receive and transmit, respectively. The relay whose timer  expires  first, broadcasts a packet
containing pilot symbols   and a control message with information  about which  relay is  selected and whether the selected relay   receives or transmits.
 From the  packet broadcasted by the selected relay,  both source and destination learn the channels from the selected relay to the source and destination, respectively. They also learn which relay is  selected  and whether it is scheduled to receive or transmit. 
If the selected relay is scheduled to transmit, then it extracts bits from its buffer, maps them to a codeword and  transmits the codeword to the destination    with rate $R_{kD}(i)=\min\{Q_k(i-1),C_{kD}(i)\}$.
Otherwise, if the relay is scheduled to receive, then  the source transmits to the selected relay a codeword with rate $C_{Sk}(i)$.

Again, the destination  may receive the information bits in an order which is different from that in which they were transmitted by the source. Therefore, in order for the destination to reorder  the received information bits, it should keep track of the amount of information  received and transmitted by each relay in each time slot. If the selected relay transmits, by successful decoding  the destination learns   the  amount of information  received. However, when the selected relay is scheduled to receive, the relay should feedback the amount of information that it received to the destination.  Using this information, the destination can perform successful reordering of the received information bits.  

\begin{remark}\label{remark_3}
 We note that distributed relay selection protocols based on timers  may suffer from long waiting times before the first timer expires. Moreover,  collisions are possible when  two or more relay nodes declare  that they are the selected node at approximately the same time. However, by choosing the timers suitably, as proposed in \cite{5474645}, these negative effects can be minimized. 
\end{remark}

\subsection{Comparison of the Overhead of the Conventional and the Proposed  Protocols}
 The conventional relay selection protocol  reviewed in Section~\ref{sec_II_conv}   can also be implemented in a centralized  or a distributed manner. In the following, we discuss the overheads entailed by both implementations.

 For the centralized implementation, the destination controls the relay selection. To this end, the destination has to acquire the CSI of all links in the network. Therefore, for centralized implementation,  in each time slot, $2M+2$ pilot symbol transmissions are required for CSI acquisition,  one control packet transmission by the destination is needed   to inform the relays  which relay is selected, and  another control packet transmission is required for the selected relay  to inform the source which  relay is selected. Moreover, the source has to acquire knowledge of  $ \min\{C_{Sk}(i),C_{kD}(i)\}$ in order to select the rate of transmission. Hence, if $\min\{C_{Sk}(i),C_{kD}(i)\}=C_{kD}(i)$, the selected relay has to feedback the CSI of the selected-relay-to-destination link to the source. As a result, in total $2M+4$ or $2M+5$ pilot symbol, feedback, and control packet transmissions are needed in each time slot. On the  other hand, for the centralized implementation of the proposed buffer-aided relaying protocol, also  $2M+4$ or $2M+5 $  pilot symbol, feedback, and control packet transmissions are required. Hence, both the conventional and the proposed buffer-aided relaying protocols have identical overheads when  implemented centrally.

For conventional relay selection  with distributed implementation, each relay has to acquire the CSI of its source-to-relay and relay-to-destination links. To this end, two pilot transmissions, one from the source and the other from the  destination, are needed. Moreover, one packet with pilots and a control message from the selected relay are needed to inform  source and destination which   relay is selected, and to allow  source and destination to learn the CSI of the source-to-selected-relay and   selected-relay-to-destination links, respectively. Furthermore, assuming relay $k$ is the selected relay in time slot $i$, in order for the 
 source to adapt its transmission rate to $ \min\{C_{Sk}(i),C_{kD}(i)\}$ and  the destination  to know which codebook to use for decoding in time slot $i$, both source and relay have to know $\min\{C_{Sk}(i),C_{kD}(i)\}$. Acquiring this CSI knowladge requires feedback    of the source-to-relay or the relay-to-destination channel from the relay to the destination or the source, respectively. Hence,  the distributed implementation of conventional relay selection  requires $4$  pilot symbol, feedback, and control packet transmissions. On the other hand, the distributed implementation of the proposed buffer-aided relaying protocol  has the same overhead as conventional relay selection since it also requires   $4$  pilot symbol, feedback, and control packet transmissions.

 As can be seen from the above discussion,  the proposed buffer-aided protocol  does not require more signaling overhead than the conventional relay selection protocol. We note, however, that the proposed protocol requires the computation of $\mu_k^e(i)$ and $Q_k(i)$, $\forall k$, which are not required for the conventional protocols. On the other hand, the computational complexity of obtaining $\mu_k^e(i)$ and $Q_k(i)$ using (\ref{eq_f1})-(\ref{eq_f2b}) and (\ref{eq_qqq1}), respectively, is not high since these equations require only  one or two additions and one to three multiplications.


\section{Buffer-Aided Relaying Protocol with a Delay Constraint}\label{sec_delay-I}

The protocol in  Theorem~\ref{theo_2}, with the $\mu_k^*$, $\forall k$, obtained from Lemma~\ref{lema_2}, gives the maximum average achievable rate, but   introduces unbounded delay. To bound the delay, in the following, we propose a buffer-aided relaying protocol for delay limited transmission.  Before presenting the protocol, we first determine the average delay for the considered network.
 
\subsection{Average Delay}\label{sec_delay}
The  average delay for the considered network, denoted by $\bar T$, is specified in the following lemma.
\begin{lemma}\label{lema_delay}
The average delay for the considered network is given by
\begin{eqnarray}\label{eq_delay}
    \bar T=\frac{\sum_{k=1}^M \bar Q_k}{\sum_{k=1}^M\bar R_{Sk}},
\end{eqnarray}
where $\bar R_{Sk}$ is the average rate  received at the $k$-th relay  and given by (\ref{eq_3s}). Furthermore, $\bar Q_k$ is the average queue size  in the buffer of the $k$-th relay, which is found as 
\begin{eqnarray}\label{eq_q_avg}
    \bar Q_k=\lim_{N\to\infty}\frac{1}{N}\sum_{i=1}^N Q_k(i).
\end{eqnarray}
\end{lemma}
\begin{IEEEproof}
Please refer to Appendix~\ref{app_delay}.
\end{IEEEproof}

The queue size at time slot $i$  can be obtained using (\ref{eq_qqq1}).
Due to the recursiveness of the expression in (\ref{eq_qqq1}), it is   difficult, if not impossible, to obtain an analytical expression for the average queue size $\bar Q_k$  for a general buffer-aided relay selection protocol. Hence, in contrast  to the case without delay constraint, for the delay limited case, it is very difficult to formulate an optimization problem for maximization of the average rate subject to some average delay constraint. As a result, in the following, we develop a simple heuristic protocol for delay limited transmission. The proposed protocol is a distributed  protocol  in the sense that the relays themselves negotiate which relay should receive or transmit in each time slot   such that the average delay constraint is satisfied.  We note that the proposed protocol  does not need any knowledge of the statistics of the channels.

\subsection{Distributed Buffer-Aided Protocol}\label{sec_delay_1}
Before presenting the proposed heuristic protocol for delay limited transmission, we first  explain the intuition behind the  protocol.

\subsubsection{Intuition Behind the Protocol}
Assume that we have a buffer-aided protocol which, when implemented in the considered network, enforces the following relation  
\begin{eqnarray}\label{eq_r1}
    \frac{\bar Q_k}{\bar R_{Sk}} = T_0, \; \forall k, 
\end{eqnarray}
i.e., the average queue length divided by the average arrival rate in the buffer at the $k$-th relay is  equal to $T_0$. If (\ref{eq_r1}) holds $\forall k$, then by  inserting (\ref{eq_r1}) into (\ref{eq_delay}), we see that the average delay of the network will be $\bar T= T_0 $. Moreover, enforcing (\ref{eq_r1}) at the $k$-th relay  requires only local knowledge, i.e., only knowledge of the average queue length and the average arrival rate at the $k$-th relay is required. Hence, this protocol can be implemented in a distributed manner.
There are many ways to enforce (\ref{eq_r1}) at the $k$-th relay. Our preferred method for enforcing (\ref{eq_r1}) is to  have the  $k$-th relay  receive and transmit when 
$  Q_k(i)/\bar R_{Sk}<T_0 $ and   $ Q_k(i)/\bar R_{Sk} >T_0$ occur, respectively.
Moreover, we prefer a protocol in which the more  $ Q_k(i)/\bar R_{Sk}$ differs from $T_0$,   the higher the chance of selecting the $k$-th relay should  be.
 In this way, $Q_k(i)/\bar R_{Sk}$ becomes a random process which exhibits  fluctuation around its mean value $T_0$, and thereby achieves  (\ref{eq_r1})  in the long run. We are now ready to present the proposed protocol.

\subsubsection{The Proposed Protocol for Delay-Limited Transmission} Let $T_0$ be the desired average delay constraint of the system.
At the beginning of time slot $i$,  source and destination transmit pilots in successive pilot time slots. This enables the relays to acquire the CSI of their respective source-to-relay and relay-to-destination channels. Using the acquired  CSI, the $k$-th relay computes $C_{Sk}(i)$ and $C_{kD}(i)$.  Next, using $C_{Sk}(i)$ and the amount of normalized information in its buffer, $Q_k(i-1)$,   the $k$-th relay computes a variable   $\lambda_k(i)$ as follows
\begin{eqnarray}\label{eq_dl1}
     \lambda_k(i)=\lambda_k(i-1)+\zeta_k(i) \left(T_0-  \frac{Q_k(i-1)}{ \bar R_{Sk}^e(i-1)}\right),
\end{eqnarray}
where $\bar R_{Sk}^e(i-1)$ is a real-time estimate  of  $\bar R_{Sk}$, computed using (\ref{eq_f2a}). In (\ref{eq_dl1}), $\zeta_k(i)$ is the step size function, which is  some properly chosen monotonically decaying function of $i$  with $\zeta_k(1)<1$.
Now, using $C_{Sk}(i)$, $C_{kD}(i)$, $Q_k(i-1)$, and $\lambda_k(i)$, the $k$-th relay turns on a timer proportional to 
\begin{equation}
    \frac{1}{\max\{\lambda_k(i) C_{Sk}(i)\;,\;\min\{Q_k(i-1),C_{kD}(i)\}/\lambda_k(i)\}}.
\end{equation}
This procedure is performed by all $M$ relays. If 
\begin{align} 
&\max\{\lambda_k(i) C_{Sk}(i)\;,\;\min\{Q_k(i-1),C_{kD}(i)\}/\lambda_k(i)\}\nonumber\\
&=\lambda_k(i) C_{Sk}(i)\nonumber
\end{align}
 and 
\begin{align} 
  & \max\{\lambda_k(i) C_{Sk}(i)\;,\;\min\{Q_k(i-1),C_{kD}(i)\}/\lambda_k(i)\}\nonumber\\
& = \min\{Q_k(i-1),C_{kD}(i)\}/\lambda_k(i), 
\end{align}
 the $k$-th relay knows that if it is selected, then it will receive and transmit, respectively. The relay whose timer  expires  first, broadcasts a control packet
containing pilot symbols   and information  about which  relay is  selected and whether the selected relay   receives or transmits. From the  packet broadcasted by the selected relay,  both source and destination learn the  source-to-selected-relay   and  the selected-relay-to-destination channels, respectively, and   which relay is  selected  and whether it is scheduled to receive or transmit. If the selected relay is scheduled to transmit, then it extracts information from its buffer and transmits a codeword to the destination with rate $R_{kD}(i)=\min\{Q_k(i-1),C_{kD}(i)\}$. However, if the relay is scheduled to receive, then the source transmits a codeword to the $k$-th relay with rate $R_{Sk}(i)=C_{Sk}(i)$. In this case, the relay has to feedback its source-to-relay   channel to the destination. This fedback CSI  is needed by the destination to keep track of the amount of information that each relay receives and transmits in each time slot so that the destination can perform successful reordering of the received information bits. Moreover, exploiting   (\ref{eq_qqq1}), this information is used by the destination to  compute the queue length in the buffer at each relay, $Q_k(i)$.

\begin{remark}
Note that with (\ref{eq_dl1}) we achieve the aforementioned goal of increasing the probability of selecting the $k$-th relay when $ Q_k(i)/\bar R_{Sk}$ differs more from  $T_0$. More precisely,   if  $ Q_k(i)/\bar R_{Sk} <T_0$, then $\lambda_k(i)$ increases and $1/\lambda_k(i)$ decreases, giving  the $k$-th relay a higher chance  to be selected for reception. On the other hand,  if  $ Q_k(i)/\bar R_{Sk}>T_0$, then $\lambda_k(i)$ decreases and $1/\lambda_k(i)$ increases, giving the $k$-th relay  a higher chance   to be selected for transmission.
\end{remark}

\begin{remark}
The required overhead of the proposed distributed delay-limited protocol is identical to the  overhead of the proposed distributed  protocol without  delay  constraint. Furthermore,  the  delay-limited buffer-aided protocol can also be implemented in a centralized manner,   similar to the scheme in Section~\ref{sec_c_i}. The centralized implementation  of the delay-limited protocol  has an overhead   identical  to the  overhead of the centralized  protocol without delay  constraint, see Section~\ref{sec_c_i}. A summary of the   overheads of conventional relay selection protocols   and  the proposed  buffer-aided (BA) relaying protocols with and without   delay constraint  is given in Table~\ref{table_1}.
\end{remark}

\begin{table*} 
\begin{center}
\caption{Number of pilot symbol, feedback, and control packet transmissions   required for the conventional and the proposed buffer-aided (BA) protocols per time slot}
\begin{tabular}{|c|c |c|}
    \hline
  & Conventional &  BA protocols with and without   delay constraint \\ \hline
   Centralized  & $2M+4$ or $2M+5 $  & $2M+4$ or $2M+5 $  \\
\hline
   Distributed  & $4$  & $4$ \\
    \hline
  \end{tabular}\label{table_1}
\end{center}
\end{table*}

\section{Numerical Examples}\label{sec_3}
We assume that all source-to-relay and relay-to-destination links are impaired by Rayleigh fading. Throughout this section, we   use the abbreviation ``BA'' to denote ``buffer-aided''.

\begin{figure}
\includegraphics[width=3.75in  ]{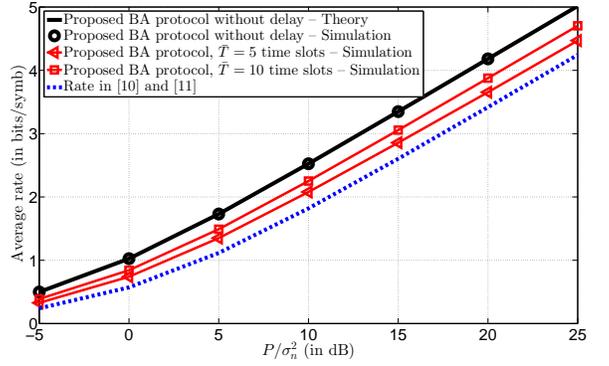}
\centering
\caption{Achievable average  rates for    $M=5$,  $[\Omega_{S1},\Omega_{S2},\Omega_{S3},\Omega_{S4},\Omega_{S5}]=[0.5,\; 1,\;1.5,\;2,\;2.5]$, and $[\Omega_{1D},\Omega_{2D},\Omega_{3D},\Omega_{4D},\Omega_{5D}]=[3,\;1.3,\;0.9,\;1.1,\;0.7]$.}\label{fig_1}
\vspace*{-3mm}
\end{figure}

\begin{figure} 
\includegraphics[width=3.75in]{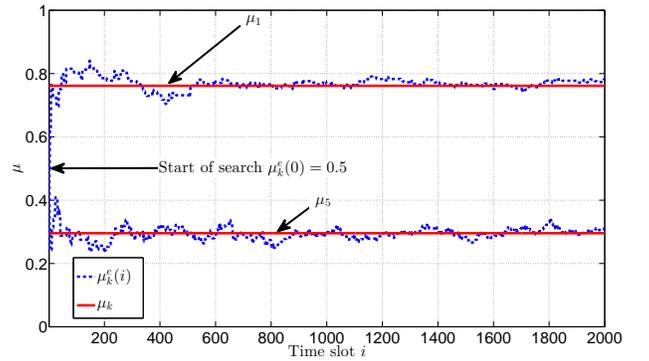}
\centering
\caption{Estimated $\mu_1^e(i)$ and $\mu_5^e(i)$  as a function of the time slot $i$.}\label{fig_1a}
\vspace*{-3mm}
\end{figure}

In Fig.~\ref{fig_1}, we   plot the theoretical maximum average rate obtained from Theorem 1, and Lemmas~\ref{lema_2} and \ref{lema_3}, for $M=5$ relays and   i.n.d.  fading, where 
\begin{eqnarray}
  [\Omega_{S1},\Omega_{S2},\Omega_{S3},\Omega_{S4},\Omega_{S5}]=[0.5,\; 1,\;1.5,\;2,\;2.5]  \nonumber
\end{eqnarray}
and
\begin{eqnarray}
\;[\Omega_{1D},\Omega_{2D},\Omega_{3D},\Omega_{4D},\Omega_{5D}]=[3,\;1.3,\;0.9,\;1.1,\;0.7].\nonumber
\end{eqnarray}
We have also included simulation results for the proposed buffer-aided protocol, where the $\mu_k^e(i)$, $k=1,...,5$, are found using the recursive method  in (\ref{eq_f1})  with   $\delta_k(i)=0.1/\sqrt{i}$, $\forall k$.   As can be seen, the simulated average rate coincides perfectly with the theoretical average rate. As a benchmark, in Fig.~\ref{fig_1}, we show the average rate  given in \cite{5351695} and \cite{5397898}.     Moreover,   we have also  included the average rates achieved using the delay limited BA protocol introduced in Section~\ref{sec_delay_1} for an average delay of $\bar T=5$ and $\bar T=10$ time slots.  For the delay limited protocol, in order to evaluate (\ref{eq_dl1})  we have used $\lambda_k(1)=0.9$ and the step size function  $\zeta_k(i)=0.005/\sqrt{i}/\log_2(1+P/\sigma_n^2)$, $\forall k$. As  can be seen from Fig.~\ref{fig_1}, both the delay-unlimited and the delay-limited BA protocols achieve higher rates than the rate  achieved   in \cite{5351695} and \cite{5397898}.
We note that we cannot use the protocols in \cite{Krikidis} and \cite{ikhlef2012max} as benchmarks in Fig.~\ref{fig_1}  since these protocols are not applicable in   i.n.d.  fading as the buffers would become unstable. In particular, for the protocols in \cite{Krikidis} and \cite{ikhlef2012max}, the buffers at relays with $\Omega_{Sk}>\Omega_{kD}$ would  suffer from overflow and receive more information than they can transmit. Hence, a fraction of the source's data would be trapped inside the buffers and does not reach the destination, i.e., data loss would occur.
 
For the parameters adopted in Fig.~\ref{fig_1},  we show in Fig.~\ref{fig_1a}   the corresponding constants $\mu_1$ and $\mu_5$ obtained using Lemma~\ref{lema_2}, and the corresponding estimated parameters $\mu_1^e(i)$ and $\mu_5^e(i)$  obtained  using the recursive method  in (\ref{eq_f1}) as   functions of   time for $P/\sigma_n^2=0$ dB. As can be seen from Fig.~\ref{fig_1a}, the estimated parameters  $\mu_1^e(i)$ and $\mu_5^e(i)$ converge relatively quickly to   $\mu_1$ and $\mu_5$, respectively.

\begin{figure} 
\includegraphics[width=3.75in ]{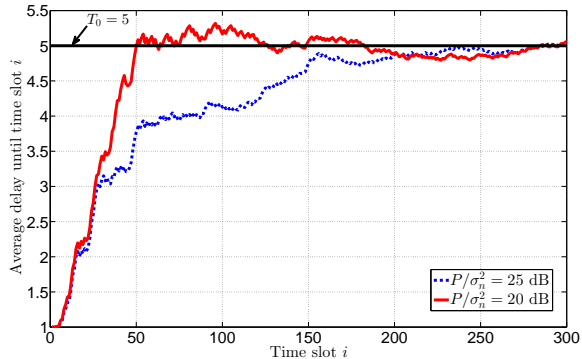}
\centering
\caption{Average delay  until time slot $i$ for $T_0=5$ and different $P/\sigma_n^2$ .}\label{fig_1b}
\vspace*{-3mm}
\end{figure}

Furthermore, for the parameters adopted in Fig.~\ref{fig_1},   we have  plotted  the average delay of the proposed delay-limited protocol until time slot $i$ in Fig.~\ref{fig_1b}, for the case when  $T_0=5$ time slots, and $P/\sigma_n^2=20$ dB and  $P/\sigma_n^2=25$ dB. The average delay until time slot $i$ is computed based on (\ref{eq_delay}) where the queue size and the arrival rates are averaged over  the time window from the first time slot to the $i$-th time slot.  Hence, for finite $i$, the average delay until time slot $i$ is the   average of a random process over a time window of  limited duration. Because of the assumed ergodicity,  for $i\to\infty$, the size of the averaging window becomes infinite and the time average converges to the mean of  this random process. However, for $i< \infty$, the time average is still a random process. This is the reason for the random fluctuations in the average delay until time slot $i$   in Fig.~\ref{fig_1b}.  Nevertheless,
 Fig.~\ref{fig_1b} shows that  the average delay until time slot $i$ converges  relatively fast to $T_0$ as $i$ increases. Moreover, after the average delay has  reached  $T_0$,  it   exhibits relatively small fluctuations around   $T_0$.

\begin{figure}
\includegraphics[width=3.75in]{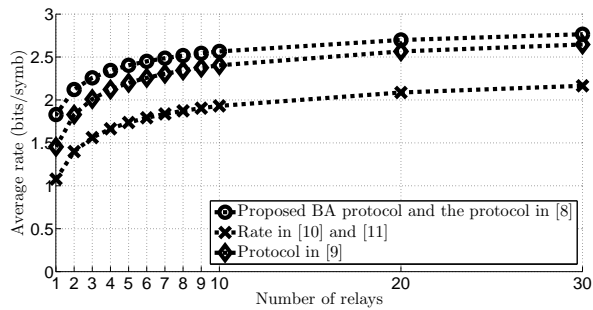}
\centering
\caption{Achievable average rates for $\Omega_{Sk}=\Omega_{kD}=1$, $\forall k$, as a function of the number of relays  $M$.}\label{fig_2}
\vspace*{-3mm}
\end{figure}

In Fig.~\ref{fig_2}, we plot the theoretical achievable average rates for  BA relaying for i.i.d. fading with $\Omega_{Sk}= \Omega_{kD}=1$, $\forall k$, and $P/\sigma_n^2=10$  dB, as a function of the number of relays $M$. As    can be seen from this numerical example, the growth rate  of the maximum average rate is inversely proportional to $M$, i.e., the growth rate  of the average data rate decreases as $M$ increases. In particular, the largest increase in data rate is observed when $M$ increases from one to two relays, whereas the increase in the maximum average rate when $M$ increases from 29 to 30 relays is almost negligible.  This behavior can be most clearly seen from the   expression for the average rate for low SNR given in (\ref{eq_d2b}). According to (\ref{eq_d2b}), the average rate increases proportionally to $1+1/2+1/3+...+1/(2M)$. Therefore, when $M$ is large, adding one more relay to the network has a negligible effect  on the average rate. 
 As   benchmarks, we also show the average rate  given in \cite{5351695} and \cite{5397898}, and the average rates achieved with the protocols in  \cite{Krikidis} and \cite{ikhlef2012max}. For i.i.d. links, as explained in Remark~\ref{remark_1}, the protocol in \cite{Krikidis} is identical to the protocol presented in Theorem~1, thereby leading to the same rate.

\begin{figure}
\includegraphics[width=3.75in]{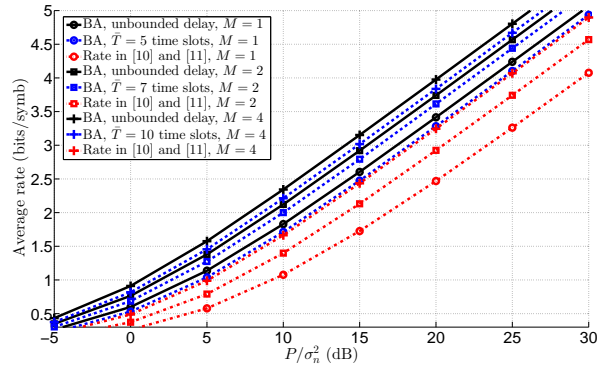}
\centering
\caption{Achievable average rates for $\Omega_{Sk}=\Omega_{kD}=1$, $\forall k$, vs $P/\sigma_n^2$ for different number of relays $M$, and different delay.}\label{fig_4}
\vspace*{-3mm}
\end{figure}

In Fig.~\ref{fig_4}, we plot the achievable average rate for  BA relaying  without and with a delay constraint,    as a function of  $P/\sigma_n^2$, for i.i.d. fading and different numbers of relays $M$. This numerical example shows that, as the number of relays increases, the permissible average delay has to be increased in order for the  rate of the delay constrained protocol to approach the rate of the non-delay constrained protocol. More precisely,  for a single relay network, an average delay of five time slots is sufficient for the rate of the delay constrained protocol to approach the rate of the non-delay constrained protocol. However, for a network with two and four relays, the corresponding required delays are $7$ and $10$ time slots, respectively.
For comparison,   we have also plotted the average rate  given in \cite{5351695} and \cite{5397898}, which requires a delay of one time slot. Fig.~\ref{fig_4} shows that the average rate of the buffer-aided relaying protocol with five time slots delay and only one relay surpasses the average  rate in \cite{5351695} and \cite{5397898}  for   four relays.  


\section{Conclusion}\label{sec_4}

We have devised buffer-aided relaying protocols for the  slow  fading HD relay selection network and derived the corresponding achievable  average rates. We have proposed a buffer-aided protocol which maximizes the achievable average rate but introduces an unbounded delay,  and a buffer-aided protocol which  bounds the average delay at the expense of a decrease in rate. 
We have shown that the new achievable rates are larger than the rates achieved with existing  relay selection protocols. We have also provided centralized and distributed implementations of the proposed buffer-aided protocols, which do  not cause more signaling overhead than conventional relay selection protocols for adaptive rate transmission  and do not need any a priori knowledge of the statistics of the involved channels.


\appendix

\subsection{Proof of Lemma~\ref{lema_1}}\label{app_lema_1}

 We denote the left and right hand sides of (\ref{eq_k}) as  $A_k$ and $D_k$, respectively, i.e.,
\begin{align}
    A_k&= \lim_{N\to\infty}\frac{1}{N}\sum_{i=1}^N r_k^{\rm R}(i) C_{Sk}(i),  \\
D_k &=\lim_{N\to\infty}\frac{1}{N}\sum_{i=1}^N r_k^{\rm T}(i) C_{kD}(i).
\end{align}
 There are three possible cases for the relationship between $A_k$ and $D_k$, i.e., $A_k>D_k$, $A_k<D_k$, and $A_k=D_k$. 
If $A_k>D_k$ then the buffer of the $k$-th relay is receiving more information than it transmits. Therefore, the average queue length in the buffer grows with time to infinity,  and, as a result,  $\bar R_{kD}=D_k$,  for a proof please refer to \cite[Section 1.5]{gross1998fundamentals}. Whereas, if $A_k<D_k$, due to the conservation of flow, the buffer cannot emit more information than it receives, and therefore  $\bar R_{kD}=A_k$. We now prove that for $A_k>D_k$ and $A_k<D_k$, $\bar R_{kD}$ can always be increased by changing the values of $r_k^{\rm R}(i)$  and $r_k^{\rm T}(i)$.  As a result, the only remaining possibility is that $\bar R_{SD}$ is maximized for  $A_k=D_k$. 
Furthermore, since the achievable rate is given by   $\bar R_{SD}=\sum_{k=1}^M \bar R_{kD}$, if $\bar R_{kD}$ increases,  $\bar R_{SD}$ will also increase.

Assume first that $A_k>D_k$. Then, we can always increase $D_k$, and thereby increase $\bar R_{kD}$, by switching any   $r_k^{\rm R}(i)=1$ for which $Q_k(i-1)>0$ holds, from one to zero and, for the same $i$, switch  $r_k^{\rm T}(i)$  from zero to one. On the other hand, if $A_k<D_k$ then we can always increase $A_k$, and thereby increase $\bar R_{kD}$, by switching any randomly chosen $r_k^{\rm T}(i)=1$ from one to zero and, for the same $i$, switch  $r_k^{\rm R}(i)$  from zero to one. Now, since $\bar R_{SD}$ can always be improved when $A_k>D_k$ or $A_k<D_k$, it follows that $\bar R_{SD}$ is maximized for  $A_k=D_k$. Furthermore, when  the $\bar R_{kD}$ are maximized  $\forall k$, then $\bar R_{SD}$ is also maximized.  Moreover, for $A_k=D_k$ the buffer at the $k$-th relay is stable since the information that arrives at the buffer also leaves the buffer without information loss. 
 On the other hand, the proof that (\ref{eq_ka}) holds when (\ref{eq_k}) is satisfied is given in \cite[Appendix B]{BA-relaying-adaptive-rate}. Finally, considering (\ref{eq_2}), if (\ref{eq_ka}) holds $\forall k$, then (\ref{eq_rate})  holds as well. This concludes the proof.

\subsection{Proof of Theorem~\ref{theo_2}}\label{app_a}

To solve (\ref{MPR1}), we first  relax  the binary constraints $r_k^{\rm T}(i)\in\{0,1\}$ and $r_k^{\rm R}(i)\in\{0,1\}$ in (\ref{MPR1})  to $0\leq r_k^{\rm T}(i)\leq 1$ and $0\leq r_k^{\rm R}(i)\leq 1$, $\forall i$, respectively.  Thereby, we transform the original problem  (\ref{MPR1}) into the following linear optimization problem
 \begin{align}\label{eq:17a}
\begin{array}{ll}
 {\underset{r_k^{\rm R}(i),r_k^{\rm T}(i),\forall i,k}{\rm{Maximize: }}}&\frac{1}{N}\sum_{i=1}^N \sum_{k=1}^M r_k^{\rm T}(i) C_{k D}(i)\\
\vspace{1mm}
{\rm{Subject\;\; to: }} &{\rm C1:}\, \frac{1}{N}\sum_{i=1}^N  r_k^{\rm R}(i) C_{S k}(i) \\
&\qquad =\frac{1}{N}\sum_{i=1}^N  r_k^{\rm T}(i) C_{kD}(i),\; \forall k \\
\vspace{1mm}
  &{\rm C2:} \, 0\leq r_k^{\rm R}(i)\leq 1,\; \forall k,i \\
\vspace{1mm}
 &{\rm C3:} \, 0\leq r_k^{\rm T}(i)\leq 1,\; \forall k,i \\
\vspace{1mm}
&{\rm C4:} \, 0\leq \sum_{k=1}^M [r_k^{\rm R}(i)+r_k^{\rm T}(i)] \leq 1,\; \forall k,i. \\
\end{array}
\end{align}
In the following, we  solve the relaxed problem  (\ref{eq:17a}) and then show that the optimal values of $r_k^{\rm T}(i)$ and $r_k^{\rm R}(i)$, $\forall i,k$ are at the boundaries, i.e., $r_k^{\rm R}(i)\in\{0,1\}$ and $r_k^{\rm T}(i)\in\{0,1\}$, $\forall i,k$. Therefore,  the solution of the relaxed problem  (\ref{eq:17a})  is also the solution to the original maximization problem in (\ref{MPR1}).

Since (\ref{eq:17a}) is a linear optimization problem, we can solve it by using the method of Lagrange multipliers. The Lagrangian function for   maximization problem  (\ref{eq:17a})  is given by
\begin{align}   \label{eq:21}
 \mathcal{L}& = 
  \sum_{k=1}^M  \frac{1}{N}\sum_{i=1}^N r_k^{\rm T}(i) C_{k D}(i) 
\nonumber\\
& - \sum_{k=1}^M \mu_k  \left( \frac{1}{N}\sum_{i=1}^N  r_k^{\rm T}(i) C_{k D}(i) -\frac{1}{N}\sum_{i=1}^N  r_k^{\rm R}(i) C_{Sk}(i) \right)\nonumber\\
& -  \sum_{k=1}^M \frac{1}{N}\sum_{i=1}^{N}  \alpha_{k}^{\rm T}(i)\left( r_k^{\rm T}(i)-1\right)+\sum_{k=1}^M \frac{1}{N}\sum_{i=1}^{N}\beta_{k}^{\rm T}(i)r_k^{\rm T}(i) \nonumber\\
& -   \sum_{k=1}^M \frac{1}{N}\sum_{i=1}^{N} \alpha_{k}^{\rm R}(i)\left( r_k^{\rm R}(i)-1\right)  +\sum_{k=1}^M \frac{1}{N}\sum_{i=1}^{N} \beta_{k}^{\rm R}(i)r_k^{\rm R}(i)\nonumber\\
& - \frac{1}{N}\sum_{i=1}^{N}\phi(i) \left( \sum_{k=1}^M [r_k^{\rm R}(i)+r_k^{\rm T}(i)]-1\right) \nonumber\\
&  +\frac{1}{N}\sum_{i=1}^{N}\lambda(i) \left( \sum_{k=1}^M [r_k^{\rm R}(i)+r_k^{\rm T}(i)]\right)  ,
\end{align}
 where $\mu_k/N$,  $\alpha_{k}^x(i)/N$, $\beta_{k}^x(i)/N$, for $x\in\{\rm R,T\}$, $\phi(i)/N$,  and $\lambda(i)/N $ are  Lagrange multipliers. These multipliers have to satisfy the following conditions.

\noindent
\textbf{1)} \: \textit{Dual feasibility condition}: The Lagrange multipliers for the inequality constraints have to be non-negative, i.e., 
\begin{align} \IEEEyesnumber \label{eq:24}
&\alpha_{k}^{\rm R}(i)\geq 0,\;\alpha_{k}^{\rm T}(i)\geq 0,\;\beta_{k}^{\rm R}(i)\geq 0, \; \beta_{k}^{\rm T}(i)\geq 0,\nonumber\\
 & \phi(i)\geq 0, \;  \lambda(i)\geq 0, \quad\forall i,k.
 \end{align} have to hold.\\
 \textbf{2)} \: \textit{Complementary slackness condition}: If an inequality is inactive, i.e., the optimal solution is in the interior of the corresponding set, the corresponding Lagrange multipliers are zero. Therefore, we obtain
 \begin{align}   \label{eq:25}
&  \alpha_{k}^{\rm R}(i)\left( r_k^{\rm R}-1\right)= 0, \quad      \alpha_{k}^{\rm T}(i)\left( r_k^{\rm T}-1\right)= 0, \quad  \forall i,k   \\
&    \beta_{k}^{\rm R}(i) r_k^{\rm R} = 0, \quad   \beta_{k}^{\rm T}(i) r_k^{\rm T} = 0, \quad  \forall i,k   \\
&  \phi(i) \left( \sum_{k=1}^M [r_k^{\rm R}(i)+r_k^{\rm T}(i)]-1\right)=0,\nonumber\\
&\lambda(i) \left( \sum_{k=1}^M [r_k^{\rm R}(i)+r_k^{\rm T}(i)]\right)=0,\quad  \forall i,k .  
 \end{align}

We now differentiate the Lagrangian function with respect to $r_n^{\rm R}(i)$ and $r_m^{\rm T}(i)$,   for $n\in\{1,...,M\}$ and $m\in\{1,...,M\}$,  and equate the results to zero, respectively. This leads to the following two equations
\begin{align}  \label{eq:73a}
\mu_n C_{Sn}(i) & =   \alpha_n^{\rm R}(i) - \beta_n^{\rm R}(i)+\phi(i)-\lambda(i)  
  \\
\left( 1-\mu_m\right)C_{mD}(i) & =  \alpha_m^{\rm T}(i) - \beta_m^{\rm T}(i)+\phi(i)-\lambda(i)   .  
\end{align}  
We first show that for the optimal solution of $r_n^{\rm R}(i)$ and $r_m^{\rm R}(i)$,  $0<r_n^{\rm R}(i)<1$ and/or $0<r_m^{\rm T}(i)<1$ cannot hold for any $n,m\in\{1,...,M\}$, and only $r_n^{\rm R}(i)\in\{0,1\}$ and $r_m^{\rm T}(i)\in\{0,1\}$ can  hold $\forall  n,m=1,...,M$. We prove this by contradiction. Assume that  $0<r_n^{\rm R}(i)<1$ and   $0<\sum_{k=1}^M[r_k^{\rm R}(i)+r_k^{\rm T}(i)]<1$. Then, according to (\ref{eq:25}), $\alpha_n^{\rm R}(i) = \beta_n^{\rm R}(i)=\phi(i)=\lambda(i)=0$ must hold. Inserting this into (\ref{eq:73a}a), we obtain 
\begin{eqnarray}  \label{eq:73b}
  \mu_n C_{Sn}(i)=0. 
\end{eqnarray} 
Since  $C_{Sn}(i)$ is an RV, (\ref{eq:73b}) can  hold  only for $\mu_n=0$. However, if we assume $\mu_n=0$, and insert $\mu_n=0$ in      (\ref{eq:73a}b) by setting $m=n$, we obtain  
\begin{eqnarray}  \label{eq:73b1}
  C_{nD}(i)= \alpha_n^{\rm T}(i) - \beta_n^{\rm T}(i).
\end{eqnarray} 
Since $C_{nD}(i)$ is a non-negative RV, and since  either $\alpha_n^{\rm T}(i)$ or  $\beta_n^{\rm T}(i)$ can be larger than zero  but not both, in order for (\ref{eq:73b1}) to hold, $\beta_n^{\rm T}(i)$ must be zero and $\alpha_n^{\rm T}(i)= C_{nD}(i)$. On the other hand, if $\beta_n^{\rm T}(i)=0$, it would mean that $r_n^{\rm T}(i)=1$. However, if $r_n^{\rm T}(i)=1$ and  $0<r_n^{\rm R}(i)<1$ hold jointly,  this would violate our starting assumption that $0<\sum_{k=1}^M[r_k^{\rm R}(i)+r_k^{\rm T}(i)]<1$ holds. Hence, $0<r_n^{\rm R}(i)<1$ and   $0<\sum_{k=1}^M[r_k^{\rm R}(i)+r_k^{\rm T}(i)]<1$ cannot hold.

Now, let us  assume that  $0<r_n^{\rm R}(i)<1$ and   $\sum_{k=1}^M[r_k^{\rm R}(i)+r_k^{\rm T}(i)]=1$. Since $r_n^{\rm R}(i)<1$, then  at least one other variable $r_k^{\rm R}(i)$ or $r_m^{\rm T}(i)$ has to be larger than zero but smaller than one, where $k\in\{1,...,M\},\;\; k\neq n$, and $m\in\{1,...,M\}$. Let us assume that this variable is  $r_k^{\rm R}(i)$, where $k\neq n$. Hence, $0<r_k^{\rm R}(i)<1$, for $k\neq n$.  Then, according to (\ref{eq:25}), $\alpha_n^{\rm R}(i) = \beta_n^{\rm R}(i)=\alpha_k^{\rm R}(i) = \beta_k^{\rm R}(i)=\lambda(i)=0$, and $\phi(i)\geq 0$ must hold. Inserting these values  in  (\ref{eq:73a}a), we obtain
\begin{eqnarray}\label{eq_a_1}
    \mu_n C_{Sn}(i) =  \phi(i) =\mu_k C_{Sk}(i) .
\end{eqnarray}
However, since $C_{Sn}(i)$ and $C_{Sk}(i)$ are independent RVs,  (\ref{eq_a_1}) cannot hold for any arbitrarily chosen $i$. On the other hand, if  we assume  that instead of $r_k^{\rm R}(i)$, the variable which is larger than one is $r_k^{\rm T}(i)$, we would have obtained that
\begin{eqnarray}\label{eq_a_2}
    \mu_n C_{Sn}(i) =  \phi(i) =(1-\mu_k) C_{kD}(i) 
\end{eqnarray}
must hold. Since (\ref{eq_a_2}) also cannot hold for any arbitrarily chosen $i$, we obtain that  $0<r_n^{\rm R}(i)<1$ and   $\sum_{k=1}^M[r_k^{\rm R}(i)+r_k^{\rm T}(i)]=1$ cannot hold. Therefore, the only other possibility is that   $r_n^{\rm R}(i)\in\{0,1\}$ must hold. 

Following the same approach as above, we can also prove  that $r_m^{\rm T}(i)\in\{0,1\}$ must hold. Moreover, due to constraint C4 in (\ref{eq:17a}), it is clear that if   $r_n^{\rm R}(i)=1$, for any $n\in\{1,...,M\}$, then  $r_k^{\rm R}(i)=0$ for all $k=1,...,M$, $k\neq n$, and    $r_m^{\rm T}(i)=0$ for all $m=1,...,M$ must hold. Similarly, if   $r_m^{\rm T}(i)=1$, for any $m\in\{1,...,M\}$, then    $r_k^{\rm T}(i)=0$ for all $k=1,...,M$, $k\neq m$, and  $r_n^{\rm R}(i)=0$ for all $n=1,...,M$  must hold.
In the following, we investigate the conditions under which $r_n^{\rm R}(i)=1$  and all other  $r_k^{\rm R}(i)=0$ for $k=1,...,M$, $k\neq n$, and all other  $r_m^{\rm T}(i)=0$ for $m=1,...,M$.

Assume $r_n^{\rm R}(i)=1$. Then, $r_k^{\rm R}(i)=0$ for $k=1,...,M$, $k\neq n$, and   $r_m^{\rm T}(i)=0$ for $m=1,...,M$ must hold.   As a result,  according to (\ref{eq:25}),  $\alpha^{\rm R}_n(i)\geq 0$, $\beta^{\rm R}_k(i)\geq 0$, $\beta^{\rm T}_m(i)\geq 0$, $\phi(i)\geq 1$, and $\beta_n^{\rm R}(i)=\alpha^{\rm R}_k(i) = \alpha^{\rm R}_m(i)=\lambda(i)=0$ must hold, for $k=1,...,M$, $k\neq n$, and     $m=1,...,M$. Inserting these variables in  (\ref{eq:73a}),  we obtain the following
\begin{eqnarray}
 \mu_n C_{Sn}(i)  &\;=\;& \alpha^{\rm R}_n(i)  +\phi(i),\label{eq_ar_1}\\
\mu_k C_{Sk}(i)  &\;=\;& -\beta^{\rm R}_k(i)  +\phi(i), \;  \forall k\neq n\label{eq_ar_2}\\
(1-\mu_m) C_{mD}(i)  &\;=\;& -\beta^{\rm T}_m(i)  +\phi(i),\;  \forall m\label{eq_ar_3}.
\end{eqnarray} 
Subtracting (\ref{eq_ar_2}) from (\ref{eq_ar_1}) and subtracting (\ref{eq_ar_3}) from (\ref{eq_ar_1}), we obtain
\begin{align}  
 \mu_n C_{Sn}(i) - \mu_k C_{Sk}(i) & =  \alpha^{\rm R}_n(i)  + \beta^{\rm R}_k(i),\; \forall k\neq n\label{eq_ar_1a}\\
\mu_n C_{Sn}(i) - (1-\mu_m) C_{mD}(i) & =  \alpha^{\rm R}_n(i)  + \beta^{\rm R}_m(i),\; \forall m. \label{eq_ar_1b}
\end{align} 
Since  $\alpha^{\rm R}_n(i)  + \beta^{\rm R}_k(i)\geq 0$ and $\alpha^{\rm R}_n(i)  + \beta^{\rm R}_m(i)\geq 0$ hold, it follows that  $r_n^{\rm R}(i)=1$ when the following holds
\begin{align}  
 & \mu_n C_{Sn}(i) > \mu_k C_{Sk}(i),\;   \forall k\neq n  \nonumber \\
&\textrm{ AND }
\mu_n C_{Sn}(i) > (1-\mu_m) C_{mD}(i),  \; \forall m .\label{eq:73g}
\end{align} 
Eq.~(\ref{eq:73g}) can be written in compact form as  
\begin{eqnarray}\label{eq:73ga}
    r_k^{\rm R}(i)=1 & \textrm{if }\;  \mu_k  C_{Sk }(i) =\max \mathcal{A}(i), 
\end{eqnarray}
where  set $\mathcal{A}(i)$ is defined in (\ref{eq_A}). Following the same approach as above, we can prove that 
\begin{eqnarray}\label{eq:73gb}
    r_k^{\rm T}(i)=1 & \textrm{if }\;  (1-\mu_k)  C_{kD }(i) =\max \mathcal{A}(i) .
\end{eqnarray}
Combining (\ref{eq:73ga}) and (\ref{eq:73gb}), we obtain (\ref{eq_1_n}). This completes the proof of Theorem~\ref{theo_2}. 

\subsection{Proof of Lemma~\ref{lema_2}}\label{app_pdf}

  The optimal $\mu_k$, $\forall k$, are found from the system of $M$ equations given in (\ref{a_11c}). Using the definition of the expected value,  (\ref{a_11c}) can be written equivalently as (\ref{a_11ca}), where the RVs  $\Gamma_{
Sk}$  and $\Gamma_{kD}$    are given by 
\begin{eqnarray}\label{rav_1}
     \Gamma_{Sk}&=&\left\{
\begin{array}{cl}
\gamma_{Sk} & \textrm{if } \mu_k  C_{Sk} =\max \mathcal{A} \\
0   & \textrm{if } \mu_k  C_{Sk} \neq \max \mathcal{A} 
\end{array}
\right.,\nonumber\\
\Gamma_{kD}&=&\left\{
\begin{array}{cl}
\gamma_{kD} & \textrm{if } (1-\mu_k)  C_{kD} =\max \mathcal{A} \\
0   & \textrm{if } (1-\mu_k)  C_{kD} \neq \max \mathcal{A}.
\end{array}
\right.
\end{eqnarray}  
Hence, to find the optimal $\mu_k$, $\forall k$, we  only have to find the PDFs of $\Gamma_{
Sk}$  and $\Gamma_{kD}$,  $f_{\Gamma_{Sk}}(x)$ and  $f_{\Gamma_{kD}}(x)$, and insert them into (\ref{a_11ca}). In the following, we first derive the PDF of  $\Gamma_{
Sk}$.
 
Using (\ref{rav_1}), we can obtain the PDF of $\Gamma_{Sk}$, $f_{\Gamma_{Sk}}(x)$, for $x>0$, as 
\begin{eqnarray}\label{eq_rav-2}
    f_{\Gamma_{Sk}}(x)= f_{\gamma_{Sk}}(x) {\rm Pr}\big\{ \mu_k  C_{Sk} =\max \mathcal{A} \big\}, \;\;x>0, 
\end{eqnarray}
where $\rm{Pr}\{\cdot\}$ denotes probability. 
Note that the distribution of $f_{\Gamma_{Sk}}(x)$ for $x=0$, is not needed for the computation of the expectations in  (\ref{a_11c}) and (\ref{a_11b}). The only unknown in (\ref{eq_rav-2}) is the probability ${\rm Pr}\big\{ \mu_k  C_{Sk} =\max \mathcal{A} \big\} $. In the following, we derive this probability. To this end, we set $\gamma_{Sk}=x$, and obtain 
\begin{align}
 & {\rm Pr}\big\{ \mu_k  C_{Sk} =\max \mathcal{A} \big\}     = {\rm Pr}\big\{ \mu_k  \log_2(1+ x)  =\max \mathcal{A} \big\} \nonumber\\
&=  \prod_{j=1,j\neq k}^M{\rm Pr}\left\{\mu_j \log_2(1+\gamma_{Sj} ) < \mu_k\log_2(1+ x)    \right\}\nonumber\\
& \times  \prod_{j=1}^M{\rm Pr}\left\{(1-\mu_j) \log_2(1+\gamma_{jD} ) < \mu_k\log_2(1+ x)    \right\}\nonumber\\
&=   \prod_{j=1,j\neq k}^M{\rm Pr} \left\{  \gamma_{Sj} <  (1+ x)^{\frac{\mu_k}{\mu_j}}-1    \right\} \nonumber\\
&\times \prod_{j=1}^M{\rm Pr}\left\{\gamma_{jD} <  (1+ x)^{\frac{\mu_k}{1-\mu_j}}-1   \right\}\nonumber\\
&=  \prod_{j=1,j\neq k}^M  F_{\gamma_{Sj}} \left( (1+ x)^{\frac{\mu_k}{\mu_j}}-1    \right) \nonumber\\
&\times \prod_{j=1}^M F_{\gamma_{jD}} \left( (1+ x)^{\frac{\mu_k}{1-\mu_j}}-1   \right), \label{eq_nn_1}
\end{align}
where $F_{\gamma_\alpha}(x)$ is the CDF of $\gamma_\alpha$, for $\alpha\in\{Sk,kD\}$. Inserting (\ref{eq_nn_1}) into (\ref{eq_rav-2}), we obtain (\ref{eq_d1a}). Following a similar procedure as above,  we obtain the distribution of $\Gamma_{kD}$ given in (\ref{eq_d1b}).

Now, assume that all  source-to-relay  and relay-to-destination  links are   i.i.d. Then,  $f_{\gamma_{Sk}}(x)= f_{\gamma_{kD}}(x)= f_{\gamma}(x)$ holds $\forall k$. Moreover, $F_{\gamma_{Sk}}(x)=F_{\gamma_{kD}}(x) = F_{\gamma}(x)$ also holds  $\forall k$. As a result, (\ref{eq_d1a}) and (\ref{eq_d1b}) can be written for $x>0$ as
\begin{align}
&f_{\Gamma_{Sk}}(x)=  f_{\gamma}(x)  F_{\gamma}\left((1+x)^{\frac{\mu_k}{1-\mu_k}}-1\right)
\nonumber\\
& \times  \prod_{\underset{j\neq k}{j=1}}^M 
F_{\gamma}\left((1+x)^{\frac{\mu_k}{\mu_j}}-1\right)
F_{\gamma}\left((1+x)^{\frac{\mu_k}{1-\mu_j}}-1\right) , \label{eq_d11a}\\
&f_{\Gamma_{kD}}(x)=  f_{\gamma}(x)  F_{\gamma}\left((1+x)^{\frac{1-\mu_k}{\mu_k}}-1\right)
\nonumber\\
& \times  \prod_{\underset{j\neq k}{j=1}}^M 
F_{\gamma}\left((1+x)^{\frac{1-\mu_k}{\mu_j}}-1\right)
F_{\gamma}\left((1+x)^{\frac{1-\mu_k}{1-\mu_j}}-1\right) . \label{eq_d11b}
\end{align}
We observe that $f_{\Gamma_{Sk}}(x)$ and $f_{\Gamma_{kD}}(x)$ in (\ref{eq_d11a}) and (\ref{eq_d11b}), respectively, are both functions of $\mu_k$ and show this explicitly by redefining them as $f_{\Gamma_{Sk}}(x,\mu_k)$ and $f_{\Gamma_{kD}}(x,\mu_k)$, respectively. Moreover, from (\ref{eq_d11a}) and (\ref{eq_d11b}) we observe that  
\begin{align} \label{eqp}
   f_{\Gamma_{kD}}(x,\mu_k)=f_{\Gamma_{Sk}}(x,1-\mu_k)  
\end{align}
holds.
If we now insert (\ref{eqp})   into  (\ref{a_11ca}), we obtain
\begin{align} \label{eqp1}
   & \int_{0}^\infty  \log_2(1+x)  f_{\Gamma_{Sk}}(x,\mu_k) dx  \nonumber\\
&=  \int_{0}^\infty  \log_2(1+x)  f_{\Gamma_{Sk}}(x,1-\mu_k) dx,\;\forall k=1,...,M.
\end{align}
Now, observe that  (\ref{eqp1})   holds   if and only if $\mu_k=1-\mu_k$, which leads to $\mu_k=1/2$. 
 This concludes the proof.

\subsection{Proof of Lemma~\ref{lema_delay}}\label{app_delay}
The average delay for a system with $M$ parallel queues is well known, and given by \cite[Eq. 11.69]{mir2006computer}. After changing the notations in \cite[Eq. 11.69]{mir2006computer} to our notations, we directly obtain (\ref{eq_delay}). 
In the following, we   give an alternative, more intuitive proof of (\ref{eq_delay}).

\begin{figure}
\includegraphics[width=3.75in ]{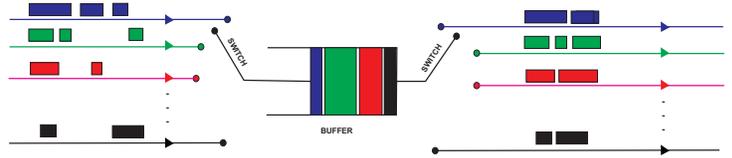}
\centering
\caption{Equivalent single buffer model.}\label{buffer_model}
\vspace*{-3mm}
\end{figure}

The input-output dynamics at the $M$ buffers in the considered network during $N$ time slots can be represented equivalently  by a   single buffer model, shown in Fig.~\ref{buffer_model}. The different colors  in this model correspond to the information bits which are received/transmitted by the different relays. For example, the blue, green, and red colors correspond to the bits that are send/received via relay 1, 2, and 3, respectively. In this model, the buffer is filled in the same order as the order of the packets that arrive at the buffers at the different relays. Which packet arrives at the equivalent buffer depends on the position of the input switch in each time slot, which on the other hand, depends on the values of $r_k^{\rm R}(i)$, $\forall i,k$. The extraction of the bits from the equivalent buffer also depends on the position of the output switch in each time slot, which on the other hand, depends on the values of $r_k^{\rm T}(i)$, $\forall i,k$. Moreover, when the  output switch is set to a line with a specific color, only  bits with that color are extracted from the equivalent  buffer. Hence, the extraction order is  different from the order of filling the equivalent buffer. Nevertheless, since the average delay computed by Little's formula \cite{little}, is independent of the order of extracting from the buffer, see  \cite[pp. 89-91]{edna_nn}, for the system model in Fig.~\ref{buffer_model}, the average delay $\bar T$ can be computed as \cite{little} 
\begin{eqnarray}
    \bar T=\frac{\bar Q_{\rm eq}}{\bar A_{\rm eq}},
\end{eqnarray}
where $\bar Q_{\rm eq}$ is the average queue size of the equivalent buffer and $\bar A_{\rm eq}$ is the average arrival rate of the equivalent buffer. Now, using the fact that $\bar Q_{\rm eq}=\sum_{k=1}^M\bar Q_k$ and $\bar A_{\rm eq}=\sum_{k=1}^M\bar R_{Sk}$, we obtain (\ref{eq_delay}). This concludes the proof.

\bibliography{litdab}
\bibliographystyle{IEEETran}

\begin{IEEEbiography}[{\includegraphics[width=1in,height=1.25in,clip,keepaspectratio]{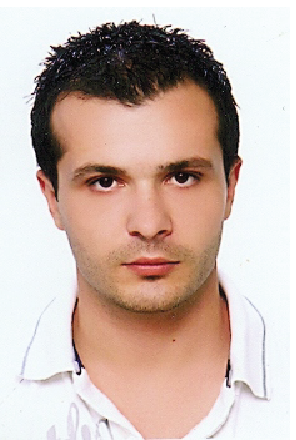}}]{Nikola Zlatanov}
(S'06) was born in Macedonia. He received his Dipl.Ing. and M.S. degrees in electrical engineering from SS. Cyril and Methodius University, Skopje, Macedonia, in 2007 and 2010, respectively. Currently, he is working toward his Ph.D. degree at the University of British Columbia (UBC), Vancouver, Canada. His current research interests include   wireless communications and information theory.

Mr. Zlatanov received several awards for his work including UBC's Four-Year Doctoral Fellowship in 2010,  UBC's Killam Doctoral Scholarship and Macedonia's Young Scientist of the Year Award in 2011, Vanier Canada Graduate Scholarship in 2012, DAAD Research Grant in 2013, and best paper award from the German Information Technology Society (ITG) in 2014.
\end{IEEEbiography}

\begin{IEEEbiography}[{\includegraphics[width=1in,height=1.25in,clip,keepaspectratio]{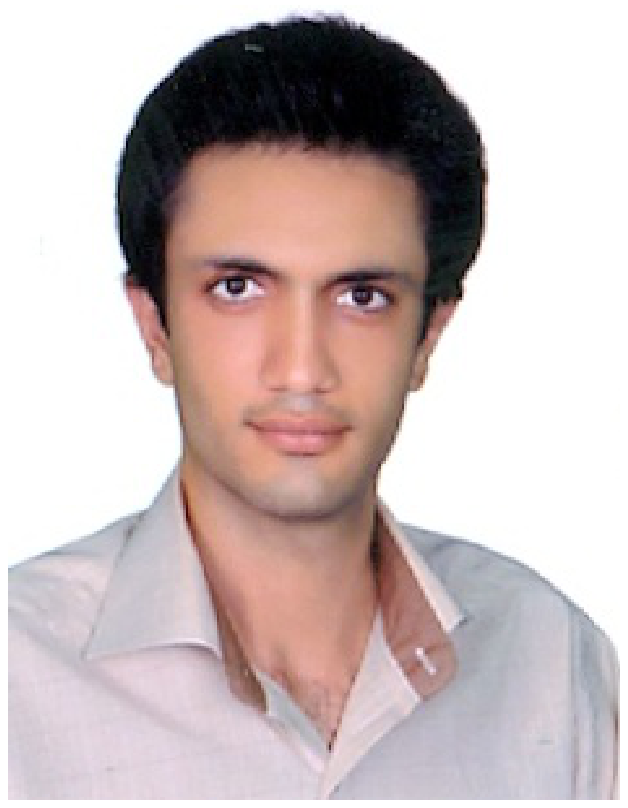}}]{Vahid Jamali} (S'12) was born in Fasa,  Iran, in 1988.  He received his B.S. and M.S. degrees in electrical engineering from K. N. Toosi University of Technology (KNTU), in 2010 and 2012, respectively. Currently, he is working toward his Ph.D. degree at the Friedrich-Alexander University (FAU), Erlangen,
Germany. His research interests include multiuser information theory, wireless communications, cognitive radio network, LDPC codes, and optimization theory.
\end{IEEEbiography}

\begin{IEEEbiography}[{\includegraphics[width=1in,height=1.25in,clip,keepaspectratio] {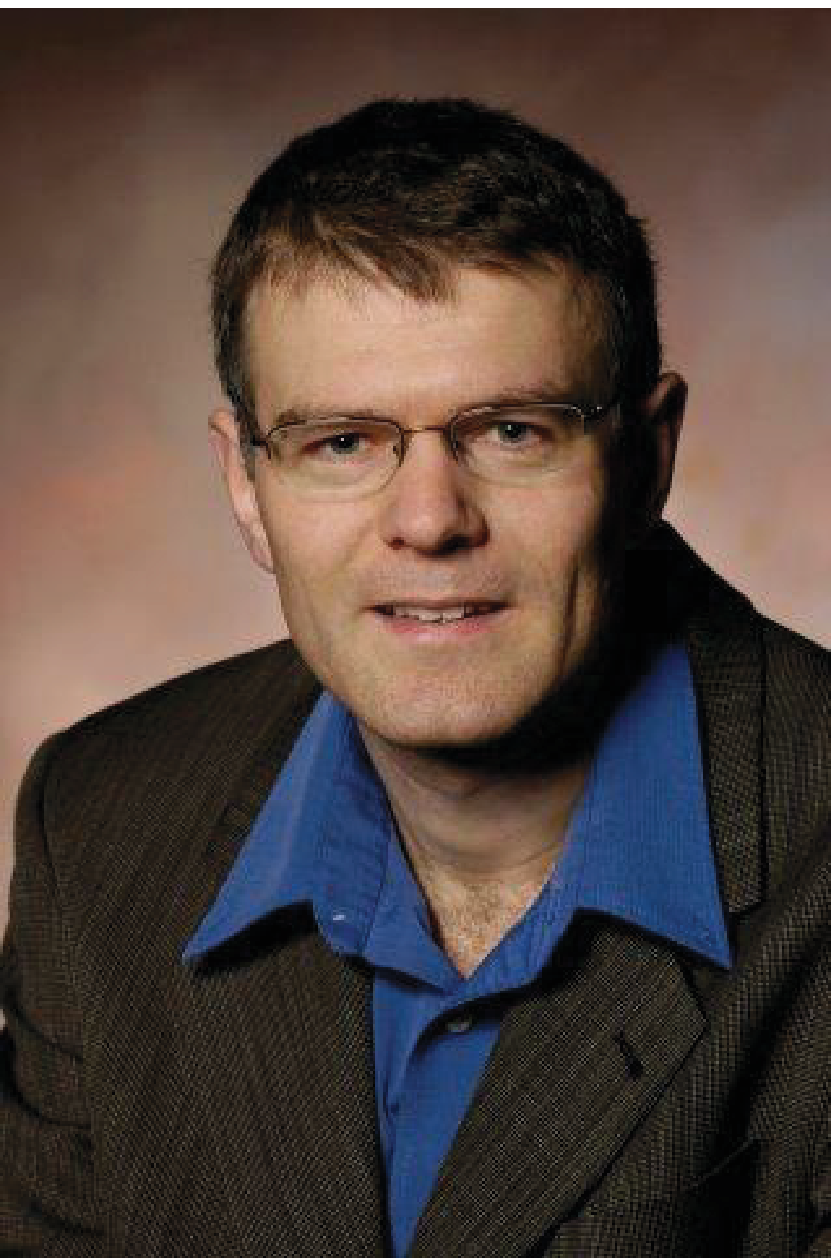}}]{Robert Schober}

 (S'98, M'01, SM'08, F'10) was born in Neuendettelsau, Germany, in 1971. He received the Diplom (Univ.) and the Ph.D. degrees in electrical engineering from the University of Erlangen-Nuermberg in 1997 and 2000, respectively. From May 2001 to April 2002 he was a Postdoctoral Fellow at the University of Toronto, Canada, sponsored by the German Academic Exchange Service (DAAD). Since May 2002 he has been with the University of British Columbia (UBC), Vancouver, Canada, where he is now a Full Professor. Since January 2012 he is an Alexander von Humboldt Professor and the Chair for Digital Communication at the Friedrich Alexander University (FAU), Erlangen, Germany. His research interests fall into the broad areas of Communication Theory, Wireless Communications, and Statistical Signal Processing.

Dr. Schober received several awards for his work including the 2002 Heinz Maier-Leibnitz Award of the German Science Foundation (DFG), the 2004 Innovations Award of the Vodafone Foundation for Research in Mobile Communications, the 2006 UBC Killam Research Prize, the 2007 Wilhelm Friedrich Bessel Research Award of the Alexander von Humboldt Foundation, the 2008 Charles McDowell Award for Excellence in Research from UBC, a 2011 Alexander von Humboldt Professorship, and a 2012 NSERC E.W.R. Steacie Fellowship. In addition, he received best paper awards from the German Information Technology Society (ITG), the European Association for Signal, Speech and Image Processing (EURASIP), IEEE WCNC 2012, IEEE Globecom 2011, IEEE ICUWB 2006, the International Zurich Seminar on Broadband Communications, and European Wireless 2000. Dr. Schober is a Fellow of the Canadian Academy of Engineering and a Fellow of the Engineering Institute of Canada. He is currently the Editor-in-Chief of the IEEE Transactions on Communications.

\end{IEEEbiography}

\end{document}